\documentclass[final,3p,12pt,times]{elsarticle}
\usepackage{graphicx}
\usepackage{color}
\usepackage{amsmath}
\usepackage{amsfonts}
\usepackage{hyperref}
\usepackage{amssymb}
\usepackage{subfigure}
\usepackage{multirow}
\usepackage{array}
\usepackage{bm}
\usepackage{wasysym}
\usepackage{threeparttable}
\usepackage[rgb,dvipsnames]{xcolor}
\definecolor{OliveGreen}{rgb}{0,0.4,0}

\journal{}

\begin{document}

\begin{frontmatter}

\title{A Multiscale Multisurface Constitutive Model for The Thermo-Plastic Behavior of Polyethylene}

\author[WM]{N. Vu-Bac}
\author[EV]{P. Areias\corref{spb}}
\author[WM,TN1]{T. Rabczuk\corref{spb}}

\cortext[spb]{Corresponding Author. Tel.: +49 (0)3643 58 4511. Email: timon.rabczuk@uni-weimar.de; pmaa@uevora.pt}
\address[WM]{Institute of Structural Mechanics, Bauhaus-Universit\"{a}t Weimar, Marienstr. 15, D-99423 Weimar, Germany}
\address[TN1]{Division of Computational Mechanics, Ton Duc Thang University, Ho Chi Minh City, Vietnam}
\address[EV]{Department of Physics, Col\'{e}gio Lu\'{i}s Ant\'{o}nio Verney, University of \'{E}vora, Rua Rom\~{a}o Ramalho, 59, 7002-554 \'{E}vora, Portugal}

\begin{abstract}

We present a multiscale model bridging length and time scales from molecular to continuum levels with the objective of predicting the yield behavior of amorphous glassy polyethylene (PE). Constitutive parameters are obtained from molecular dynamics (MD) simulations, decreasing the requirement for ad-hoc experiments. Consequently, we achieve: (1) the identification of multisurface yield functions; (2) the high strain rate involved in MD simulations is upscaled to continuum via quasi-static simulations. Validation demonstrates that the entire multisurface yield functions can be scaled to quasi-static rates where the yield stresses are possibly predicted by a proposed scaling law; (3) a hierarchical multiscale model is constructed to predict temperature and strain rate dependent yield strength of the PE.

\end{abstract}


\end{frontmatter}

\section{Introduction} 
\label{sec:introduction}

Polymers are extensively used in industrial applications, particularly in the aerospace and automotive industry due to their  physical and mechanical properties. Polymeric composites are highly complex and their mechanical properties depend on many variables such as temperature, strain rates, etc. and the physics of plastic deformations in amorphous polymers has not been well understood. Attempts were made to understand, especially, temperature and strain rate dependent yielding in polymers, see \cite{Mayr:1998,Cook:1998} for a concise review. Nevertheless, these models tend to rely on experiments whose constitutive parameters are not physically motivated and can only be used to predict behavior of a specific material. In engineering practice, a visco-plastic model based on the pressure-modified von- Mises criterion is commonly used for thermoplastic polymers \cite{Rottler:2001,Vogler:2007}. However, these polymers behave differently under tensile, compressive and shear deformations. Hence, the von-Mises yield criterion is no longer appropriate. A number of theoretical studies have been done to find proper yield surface for the prediction of thermoplastics \cite{Vogler:2007,Kolupaev:2007}. Multisurface yield functions seem a suitable candidate to describe the yield behavior for a wide range of polymers \cite{Vogler:2007}. However, calibration (fitting) procedures are not always possible for multiaxial loading conditions especially, due to loss of data.

To accurately predict macroscopic properties the molecular feature associated with the plastic mechanism must be understood \cite{Bouvard:2009}. Molecular theories of plastic behavior in amorphous polymers were reviewed by Stachurski \cite{Stachurski:1997}. However, the behavior at nano length scales was not explained within scope of these theories.

Along with the development of accurate inter-atomic potentials using quantum mechanics, molecular dynamics (MD) simulations are a powerful tool in visualizing molecular mechanisms of yielding \cite{Sundararaghavan:2013}. MD simulations offer a promising way to develop new theories and models for glassy polymers as they can reduce the need for ad-hoc experiments.

Fully atomistic models based on force fields and chemical structure of materials allow us to physically interpret their complex physical phenomena \cite{LiLiu:2012}, the length and time scales, nevertheless, limit the mechanism associated with viscoelastic/plastic behavior of the material, since interactions between single atoms are explicitly considered. Coarse graining methods, such as united atom (UA) models, can increase the length scales though the time scale has been still limited \cite{Bouvard:2009}. Rottler et al. \cite{Rottler:2001} studied shear yielding of glassy polymers under multiaxial loading conditions using MD simulations and relate them to the pressure-modified von-Mises criterion, but just at nanoscale as the simulation time is prohibitive.

However, macroscopic continuum mechanics models can be employed to study large domains and realistic time-scales. Therefore, a multiscale model passing the nanoscale descriptions to the continuum is very important. In other words, macroscopic constitutive parameters describing the evolution of macroscopic properties can be obtained from MD simulations \cite{NamLiu:2015}.

Due to high strain rates involved in MD simulations, which are not experimentally encountered, an appropriate scaling law for the yield surface \cite{NamLiu:2015} is essential to reconcile the different strain rates of MD simulations and experiments. In this article, the quasi-static simulations are employed to extract yield stresses at quasi-static strain rates from MD simulations. Furthermore, Bayes' theorem is used to construct an upscaling technique. In particular, Bayesian approach considering \emph{prior information} of the parameters (i.e. strain rate) upon which the \emph{posterior distribution} is updated given a set of observations, leading to an identification of constitutive parameters.

The article begins with the nanoscale model of the PE. Temperature dependence of the elastic and yield behavior is subsequently accounted for. Also, the Bayesian updating used to study the strain rate scaling laws is briefly depicted. The following Section describes the macroscopic continuum model whose constitutive properties are obtained from nanoscale model. Numerical results will be presented before we conclude with a discussion in Section \ref{sec:conclusion}.

\section{Nanoscale model} \label{sec:nanoscale}

\subsection{Model system and simulations} \label{subsec:model_system}

The material is described by a united atom (UA) model using the DREIDING force field \cite{Mayo:1990} with harmonic covalent potential functions and the truncated Lennard-Jones (LJ) 6-12 for non-bonded van der Waals interactions whose parameters are adopted from \cite{NamRabczuk:2012}. The functional form and parameters are presented in Table \ref{ta:dreiding_force_field}.

\begin{table}[!htb]
	\centering \caption{Functional form and parameters of the Dreiding force field}\label{ta:dreiding_force_field}
	\scalebox{0.9}{
		\newcommand*{\TitleParbox}[1]{\parbox[c]{6cm}{\raggedright #1}}%
		\begin{tabular}{l l l}
			\hline
			Interaction              & Form 	& Parameters  \\
			\hline
			Bond 		& $E_b = \frac{1}{2} k_b({r-r_{eq}})^2$									& \TitleParbox{$k_b=350~kcal/mol {\AA{}}^2,~r_{eq}=1.53~{\AA{}}^2$}  \\
			Angle           & $E_a = \frac{1}{2} k_{\theta}(cos(\theta)-cos(\theta_{eq}))^2$	& \TitleParbox{$k_{\theta} = 60~kcal/mol/rad^2,~\theta_{eq} =109.5^0$}  \\
			Dihedral	& $E_d = \frac{1}{2} \sum_{i=0}^{3} d_i cos^i (\phi)$					& \TitleParbox{$d_0 = 1.736,~d_1 = -4.490,~d_2 = 0.776,~d_3 = 6.990~(kcal/mol)$} \\
			Non-bonded     	& $E_{nb} = \left \{\begin{array}{l l}
			4 \xi \left[ \left( \frac{\delta}{r} \right)^{12} - \left( \frac{\delta}{r} \right)^{6} \right] & \quad r \leq r_{cut}  \\
			0 & \quad r > r_{cut}
			\end{array} \right.$
			
			& \TitleParbox{$\xi = 0.112~kcal/mol, \delta = 4.01~\AA{}, r_{cut} = 10.5~\AA{}$} \\
			\hline
		\end{tabular}}
\end{table}

\begin{figure}[htbp] \centering
	\subfigure[]{\includegraphics[width = 0.4\textwidth]{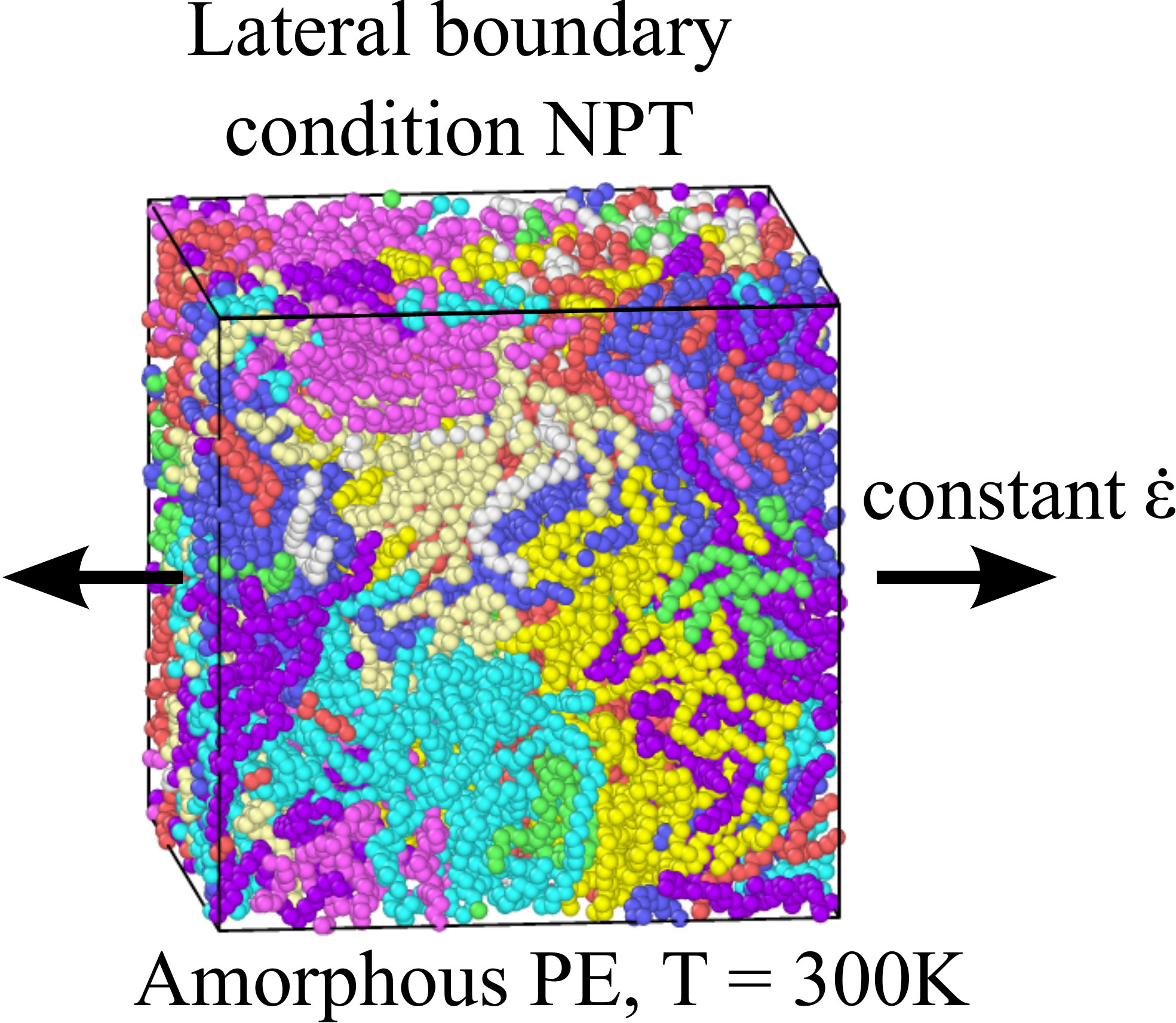}}
	\subfigure[]{\includegraphics[width = 0.45\textwidth]{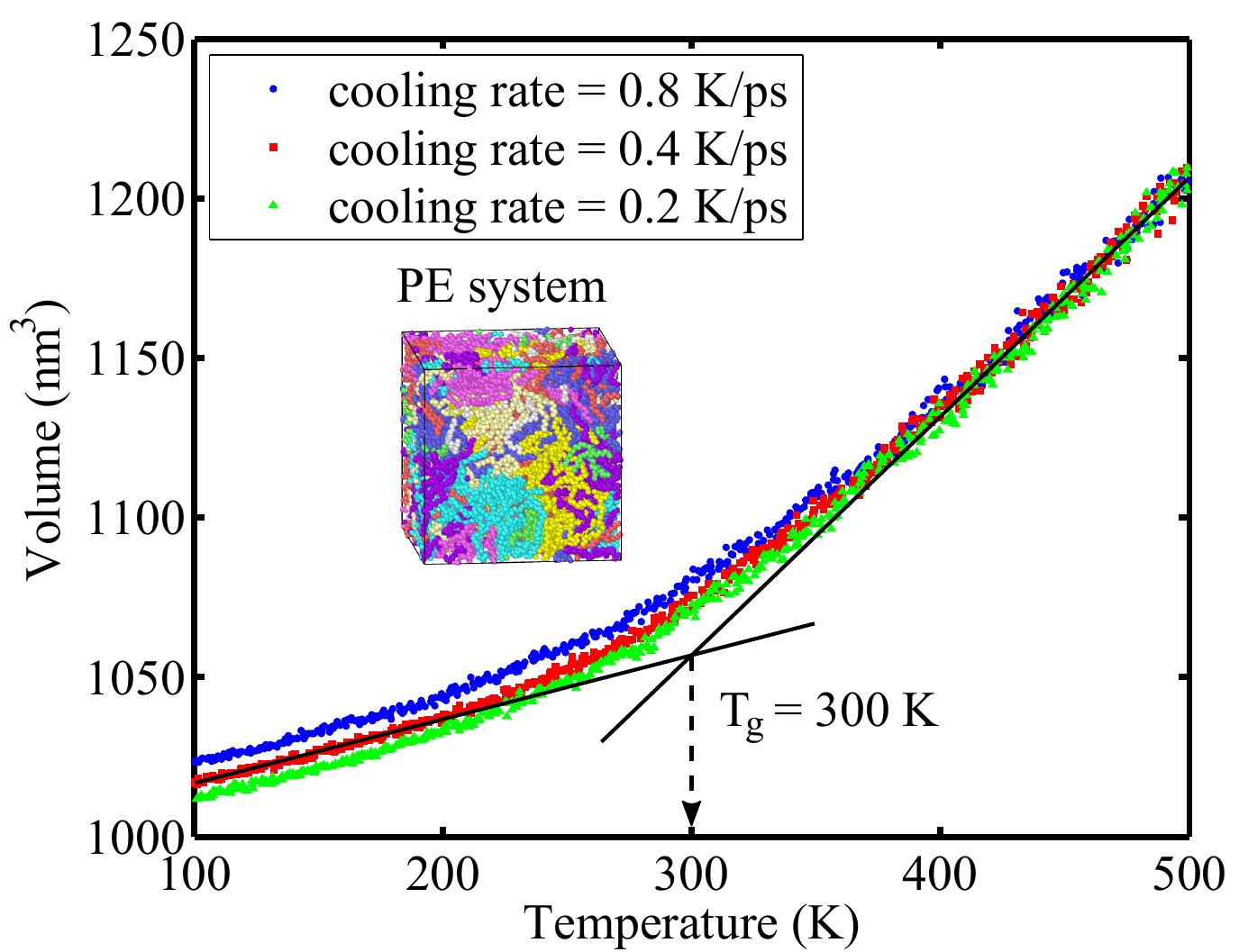}}
	\subfigure[]{\includegraphics[width = 0.45\textwidth]{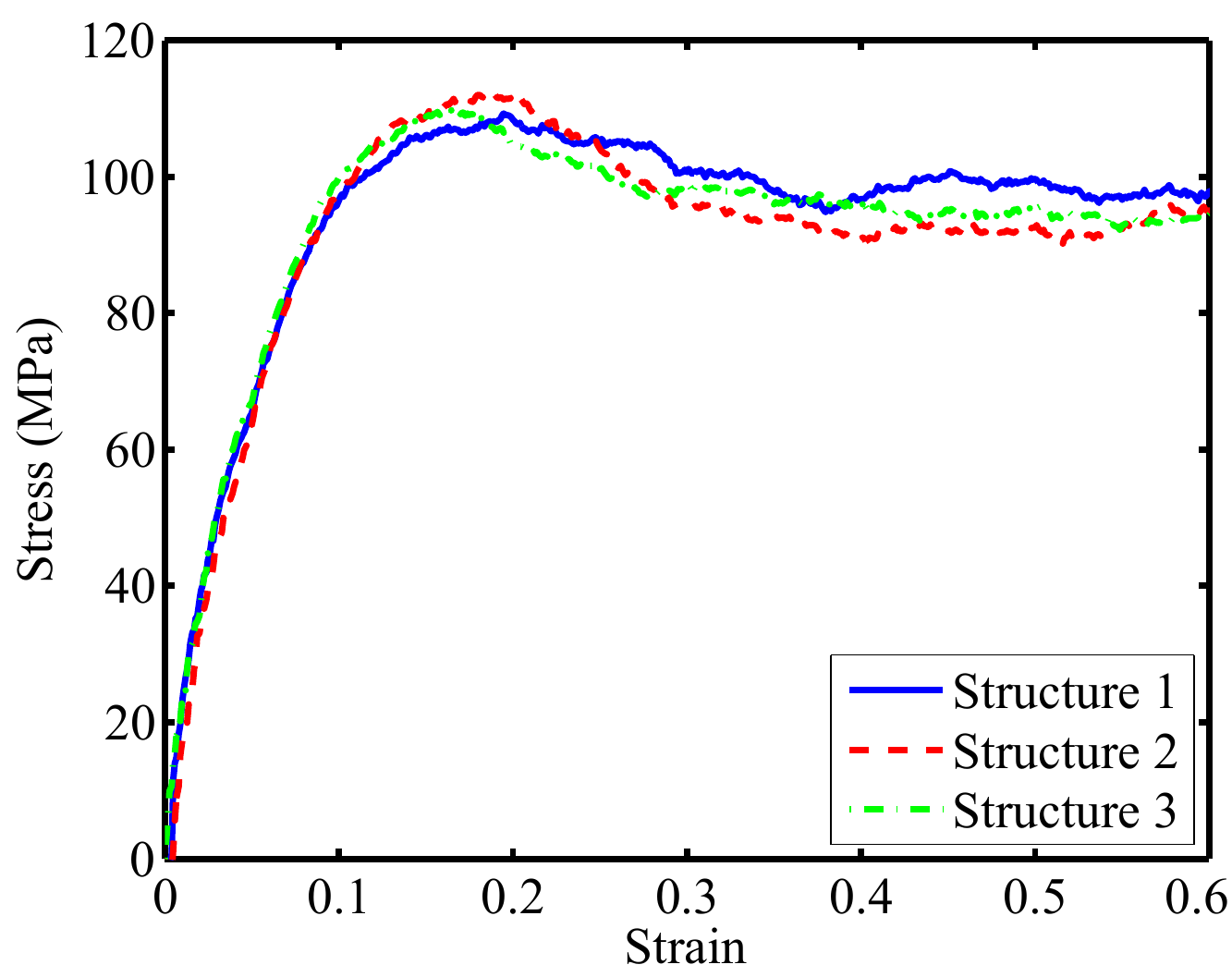}}
	\caption{(a) Undeformed system of polyethylene (PE) and boundary conditions, (b) plot of volume as a function of temperature and the glass transition temperature ($T_g$) for different cooling rates and (c) stress-strain response for different aging times at strain rate of $10^{10} ~s^{-1}$ and temperature of $300$ K: structure 1, structure 2 and structure 3 are equilibrated by $500$ ps, $1000$ ps and $2000$ ps after the cooling process, respectively.}
	\label{fig:PE_structure}
\end{figure}

The initial polymer structure was generated by using a Monte Carlo self-avoiding random walks algorithm as described by Binder \cite{Binder:1995}. A face-centered cubic (FCC) is used when generating initial configuration within a simulation box. Molecules were added to the lattice in a step-wise manner based on a method to make the appropriate selection of neighboring lattice sites. For each polymer chain, the first atom is added to an available site on the lattice. Then, the polymer chain is grown in certain directions on the bond angle and the density of the region where sites are not occupied in the probability context. LAMMPS \cite{Plimpton:1995} is employed to equilibrate the PE system through four sequential steps: (1) the PE structure was equilibrated for $10^5$ timesteps ($\Delta t = 1$fs) at $500$K using a Nose-Hoover thermostat ($NVT$) \cite{Nose:1984,Hoover:1985}; (2) a Nose-Hoover barostat ($NPT$) at the temperature of $500$K and the pressure of $1$atm condition was conducted for $5 \times 10^5$ timesteps ($\Delta t = 1$fs); (3) the structure was then cooled down to the desired temperature with a cooling rate of $0.4$K/ps followed by further $5 \times 10^5$ timesteps ($\Delta t = 1$fs) where the structure is in equilibrium. During the cooling process, the glass transition temperature ($T_g$) is determined as the intersection of two linear fitted lines to the volume versus temperature curve, see Figure \ref{fig:PE_structure}(b). Three cooling rates $0.8$ K/ps, $0.4$ K/ps and $0.2$ K/ps are used herein to take the effect of cooling rate on the glass transition temperature ($T_g$) into consideration. As observed, volume-temperature plots and the resultant $T_g$ corresponding with various cooling rates are almost identical. It is shown that $T_g = 300$K and density $\rho \approx 0.87 \div 0.91$ g/cm$^3$ are in good agreement with previous simulation and experiment results ($T_g = 250 K ~\text{and} ~\rho = 0.95 g/cm^3$ are experimentally measured value) on high density polyethylene (HDPE), see \cite{Capaldi:2004,Hossain:2010,Brandrup:1989}. Furthermore, the influence of aging time on the stress-strain response was studied where the tensile stress-strain curves deformed at strain rate of $10^{10} ~s^{-1}$ and temperature of $300$ K for three different polymer structures which are equilibrated by $500$ ps, $1000$ ps and $2000$ ps after the cooling process, respectively, are illustrated. As shown in Figure \ref{fig:PE_structure}(c), in MD simulations when polymer systems are equilibrated long enough, the ageing time insignificantly influence on the stress-strain response as the curves are nearly the same for the initial stages. Deformation simulations will be described in the sequel.

\subsection{Deformation simulations} \label{subsec:deformations_simulation}

In order to study the yield behavior of PE, the PE system was loaded in uni- and biaxially tensile/compressive strains at constant strain rates along the deformed directions. The pressure on the remaining two (uniaxial strain) or one (biaxial strain) lateral surfaces is maintained at $1$atm under \emph{NPT} dynamics. As proposed previously \cite{Bauwens:1969,Bowden:1972}, the yield stress was taken as the maximum of stress-strain responses.

\begin{figure}[htbp] \centering
	\subfigure[]{\includegraphics[width = 0.45\textwidth]{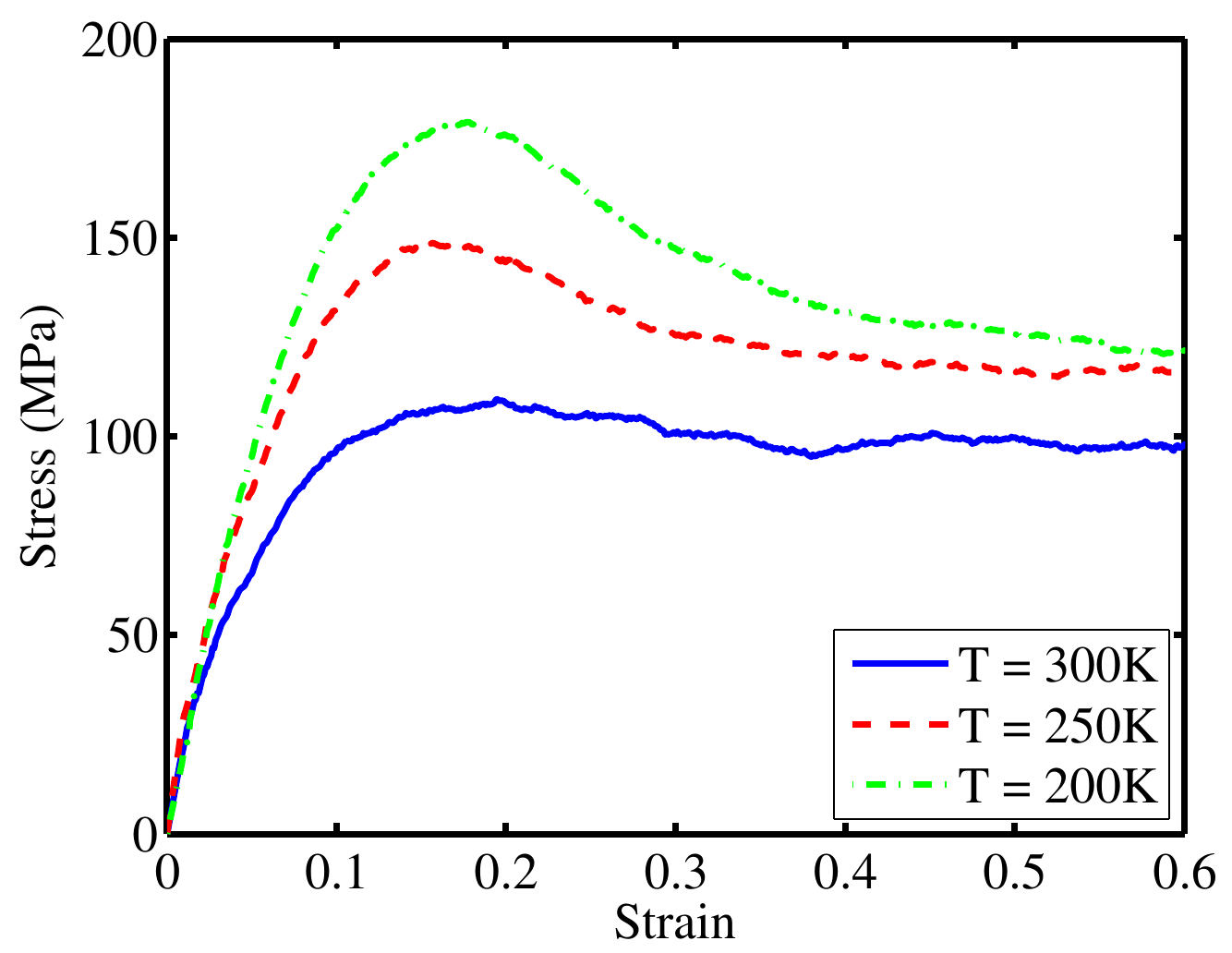}}
	\subfigure[]{\includegraphics[width = 0.45\textwidth]{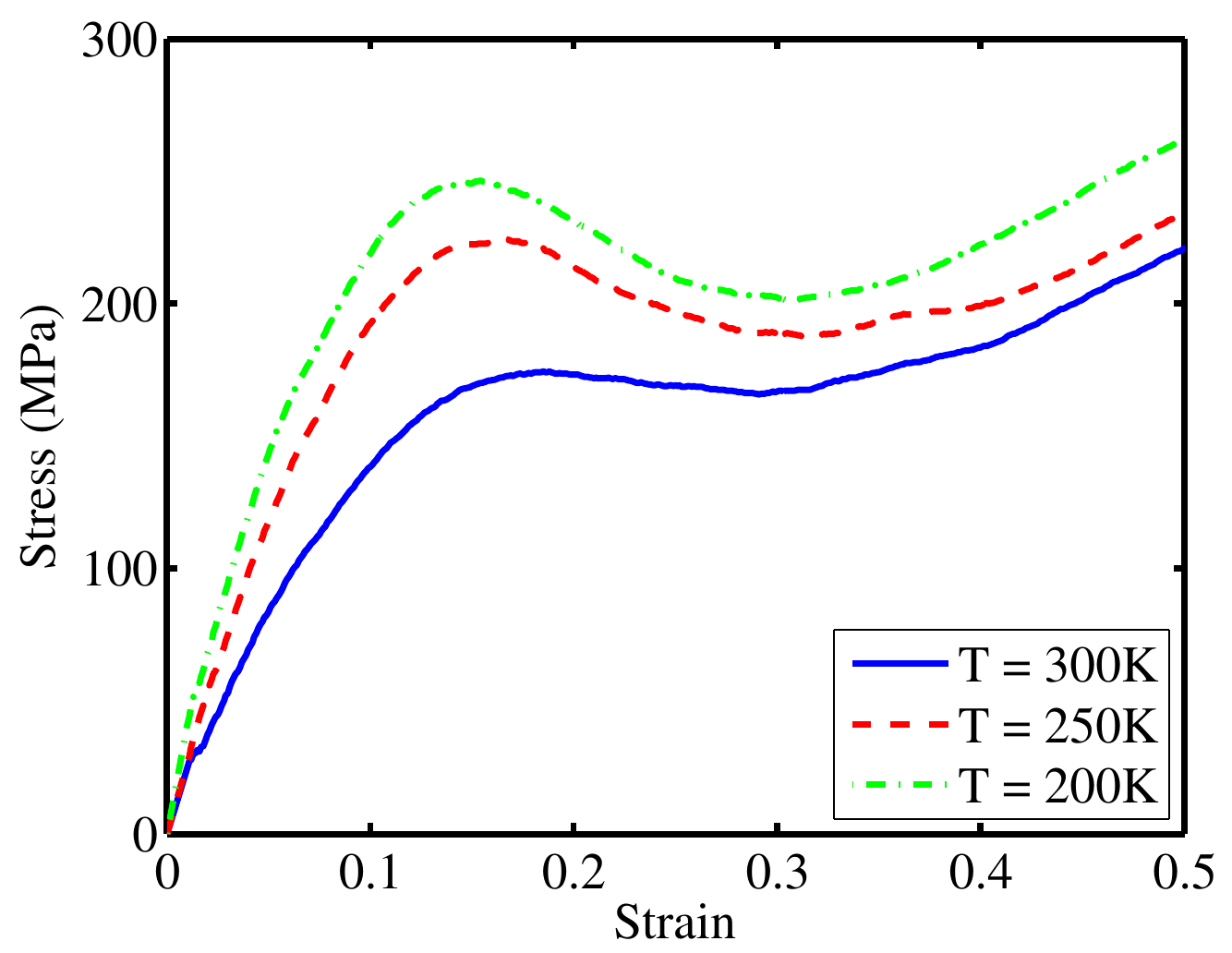}}
	\subfigure[]{\includegraphics[width = 0.45\textwidth]{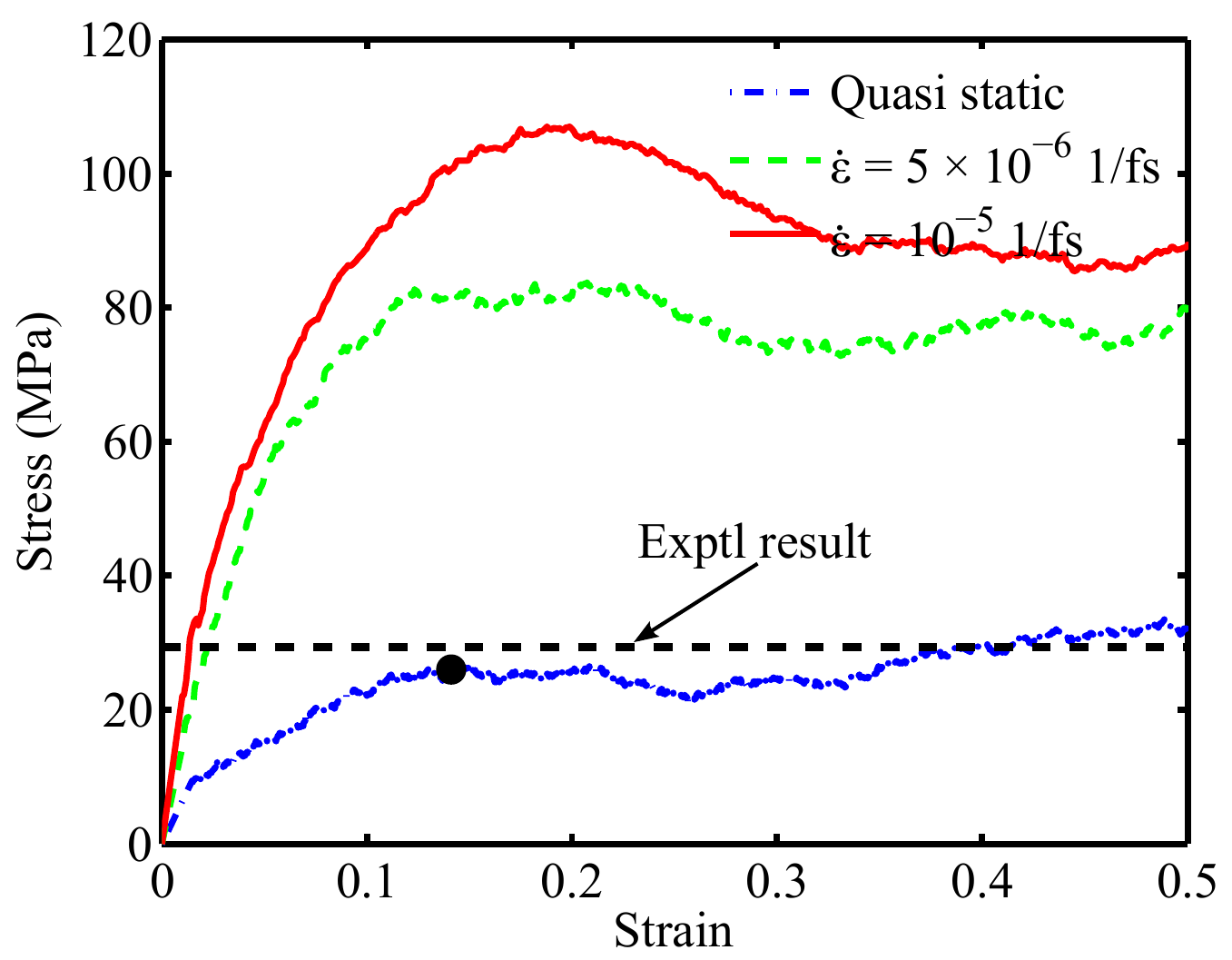}}
	\subfigure[]{\includegraphics[width = 0.45\textwidth]{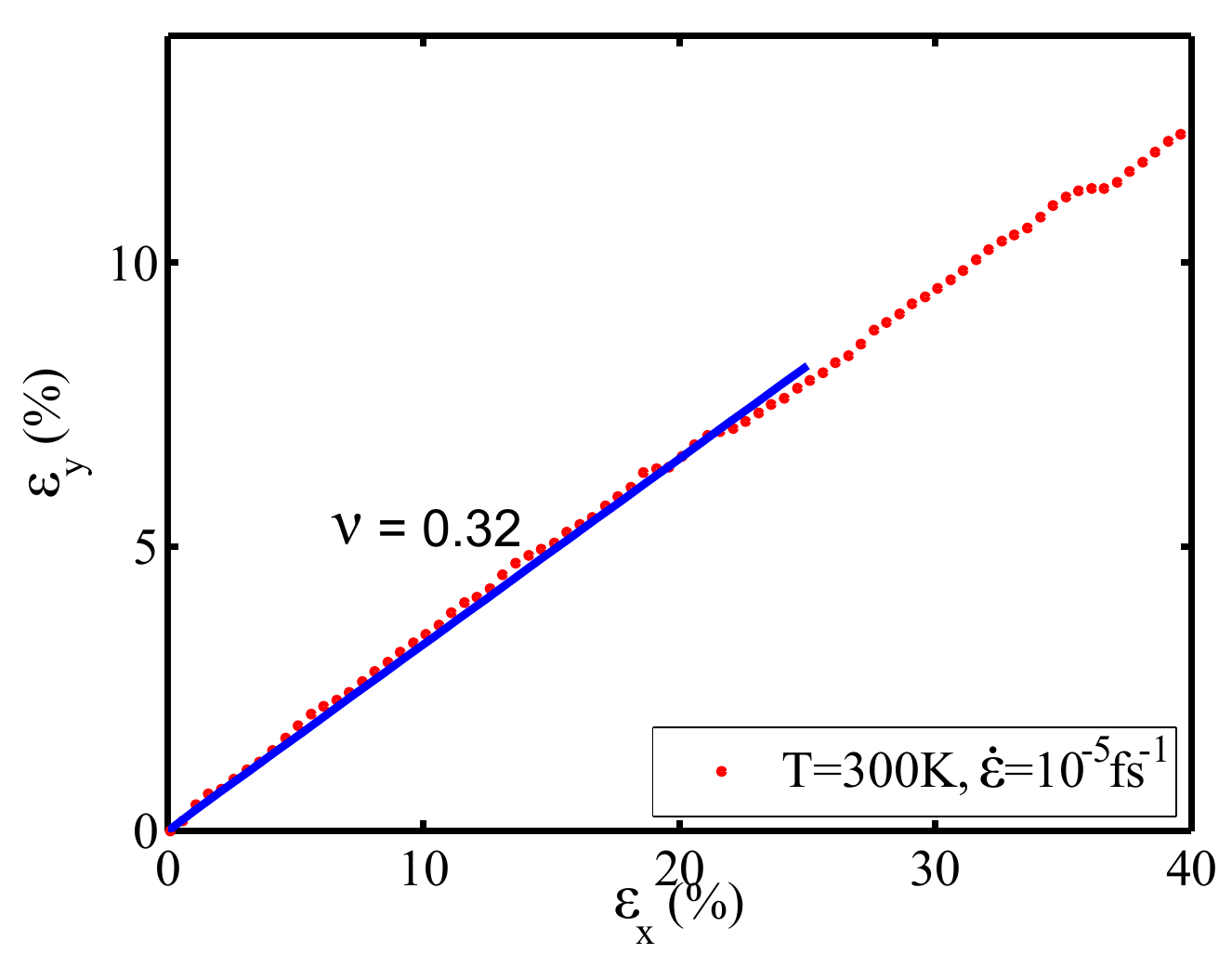}}
	\caption{Stress-strain responses under uniaxial (a) tension, (b) compression at strain rate of $10^{10} ~s^{-1}$ for different temperatures, (c) quasi-static and dynamics simulations in tension, and (d) tensile Poisson's ratio. The quasi-static tensile yield stress is indicated by ($\bullet$).}
	\label{fig:stress_strain_responses}
\end{figure}

The Young's modulus obtained from uniaxial tension at room temperature ($300$K) is $1.32$GPa, and the Poisson's ratio is $0.32$, see Figure \ref{fig:stress_strain_responses}. These results are in good agreement with experimental results: Young's modulus $E = 1.38$GPa (obtained from testing method ASTM D368) and Poisson's ratio $\nu = 0.3$ \cite{hdpe,Hartmann:1986}. Note that the mechanical properties are averaged for three different initial PE structures to take entropic effects into account as suggested by \cite{Hossain:2010}. Furthermore, the quasi-static tensile stress-strain response is simulated by using MD simulations as proposed by Capaldi et al. \cite{Capaldi:2004}. The system was uniaxially stretched at a constant strain rate of $10^{9} ~s^{-1}$ for $1000$ steps followed by equilibration for $10000$ steps ($\Delta t = 1$fs) with the axial dimension kept fixed to stabilize the energy in the system. This process is iterated until the desired strain is obtained. It is shown in Figure \ref{fig:stress_strain_responses}(c) that the quasi-static tensile yield stress ($\bullet$) is in a good agreement with experimental result reported by \cite{Hartmann:1986}.

\begin{figure}[htbp]
	\centering
	\subfigure[]{\includegraphics[width = 0.4\textwidth]{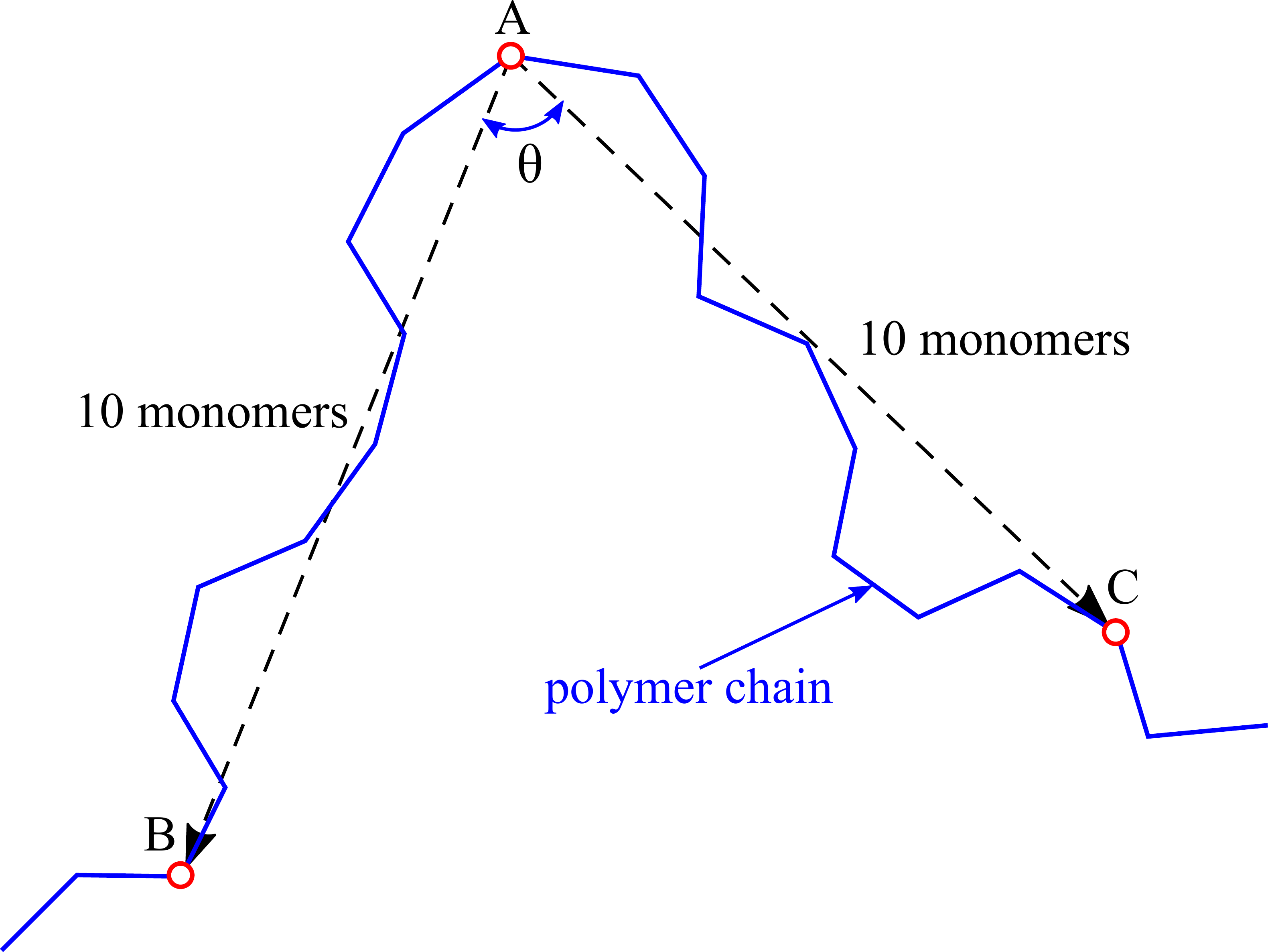}} \\
	\subfigure[]{\includegraphics[width = 0.45\textwidth]{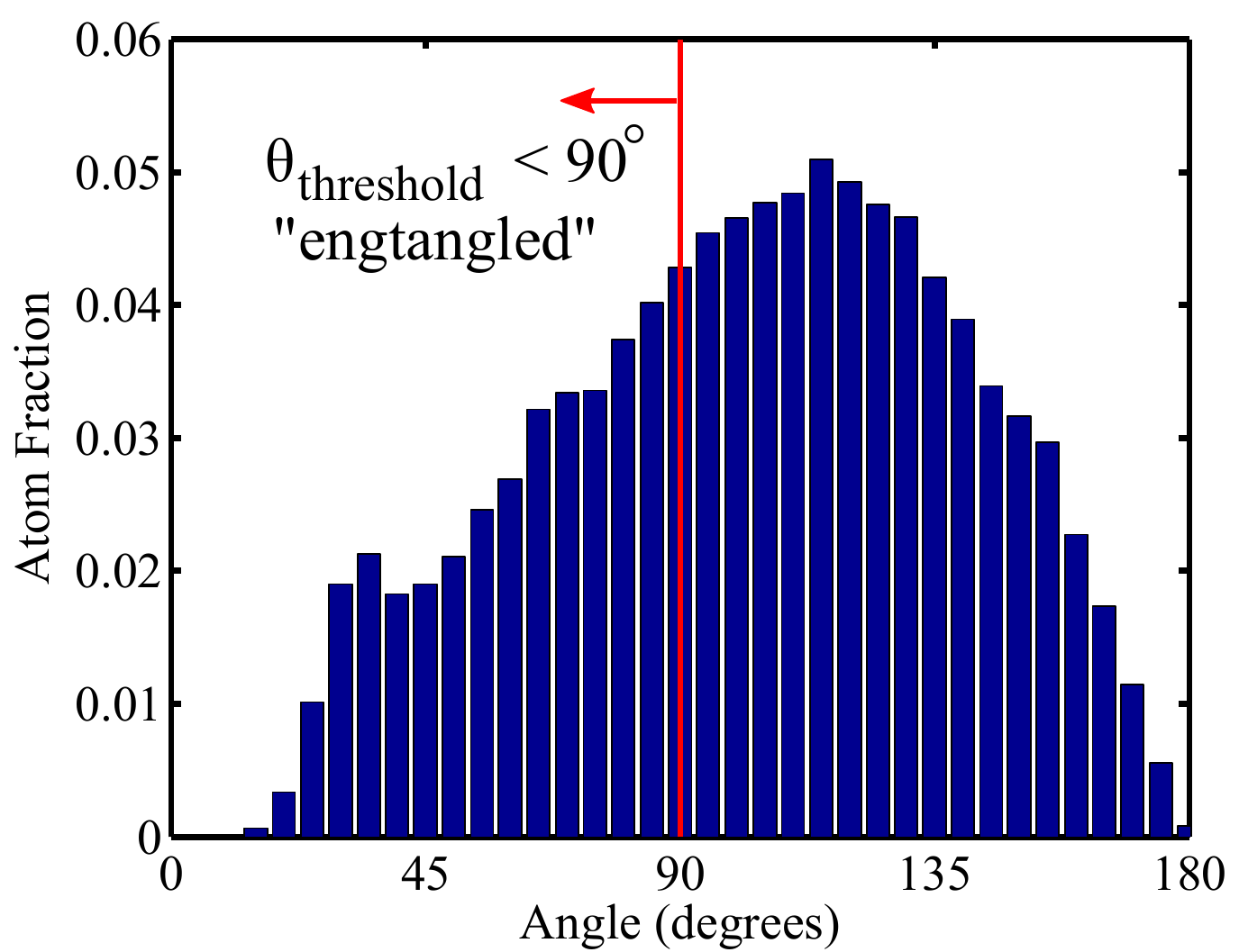}}
	\subfigure[]{\includegraphics[width = 0.45\textwidth]{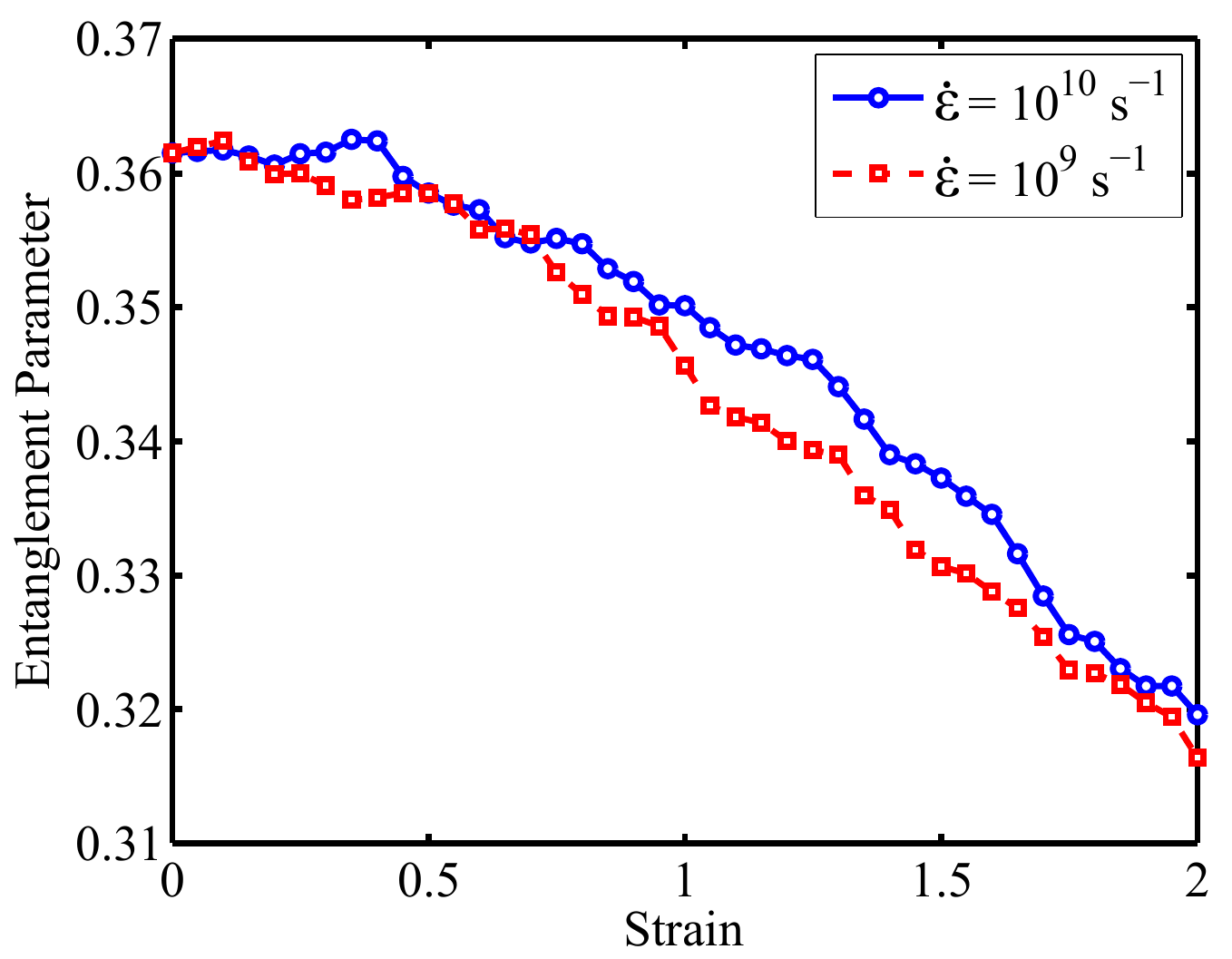}}
	\caption{(a) Schematic of technique used to estimate flexion node \cite{Yashiro:2003}, (b) Histogram of distribution of the angles estimated by the flexion node method at $250$ K and (c) plot of entanglement parameter as a function of strain for different strain rates.}
	\label{fig:entanglement}
\end{figure}

As the chain entanglement evolution is considered as important information that affects the deformation mechanisms of polymer, we have studied the chain entanglement evolution by using the geometric technique presented by Yashiro et al. \cite{Yashiro:2003}. As illustrated in Fig. \ref{fig:entanglement}(a), the interior angle $\theta$ between two vectors, i.e. one vector that is drawn from atom $i$ (A) to atom $(i-10)$ (B) and the other one that is drawn from atom $i$ (A) to atom $(i+10)$ (C), is measured. An example histogram of the distribution of the angles is shown in Fig. \ref{fig:entanglement}(b). The atoms, at which the angle $\theta$ is less than $90^0$, are classified as entangled or flexion nodes as indicated by \cite{Hossain:2010}. Furthermore, the evolution of the entanglement parameter, which is obtained by dividing the number of atoms classified as entangled by the total number of applicable atoms, as a function of strain is plotted in Fig. \ref{fig:entanglement}(c). As can be seen, the entanglement parameter, which represents the percent of entangled atoms within the system, is nearly constant for the initial stages of deformation. At lager deformation ($\varepsilon \approx 0.5$), the entanglement parameter decreases nearly linearly with an increase in strain. These results are in good agreement with previous results reported by \cite{Hossain:2010}.

\subsection{Evaluation of yield stress in multiaxial stress states} \label{subsec:multi_stress_states}

The principal stress components $\sigma_i$ are extracted from biaxially tensile and compressive deformations as proposed by \cite{Rottler:2001}. Note that the stresses obtained from biaxial loadings have to be plotted versus the equivalent strain $\bm{\varepsilon}_e$ defined by \cite{Simo:2000}:


\begin{equation} \label{eq:equivalent_strain}
	\bm{\varepsilon}_e = \sqrt{\frac{1}{2}} \sqrt{(\varepsilon_{11} - \varepsilon_{22})^2+ (\varepsilon_{22} - \varepsilon_{33})^2 + (\varepsilon_{11} - \varepsilon_{33})^2 + \frac{4}{3} \gamma^2_{12} + \frac{4}{3} \gamma^2_{23} + \frac{4}{3} \gamma^2_{13}},
\end{equation}

\noindent where $\varepsilon_{ii} ~\text{and} ~\gamma_{ij}, i,j = 1,...,3$ are three normal and shear components of the strain tensor. The equivalent strain rate applied to the PE system is provided by Equation \ref{eq:equivalent_strain}.

\begin{equation} \label{eq:equivalent_rate}
	\bm{\dot{\varepsilon}}_e = \sqrt{\frac{1}{2}} \sqrt{(\dot{\varepsilon}_{11} - \dot{\varepsilon}_{22})^2+ (\dot{\varepsilon}_{22} - \dot{\varepsilon}_{33})^2 + (\dot{\varepsilon}_{11} - \dot{\varepsilon}_{33})^2 + \frac{4}{3} \dot{\gamma}^2_{12} + \frac{4}{3} \dot{\gamma}^2_{23} + \frac{4}{3} \dot{\gamma}^2_{13}},
\end{equation}

\begin{figure}[htbp] \centering
	\subfigure[]{\includegraphics[width = 0.45\textwidth]{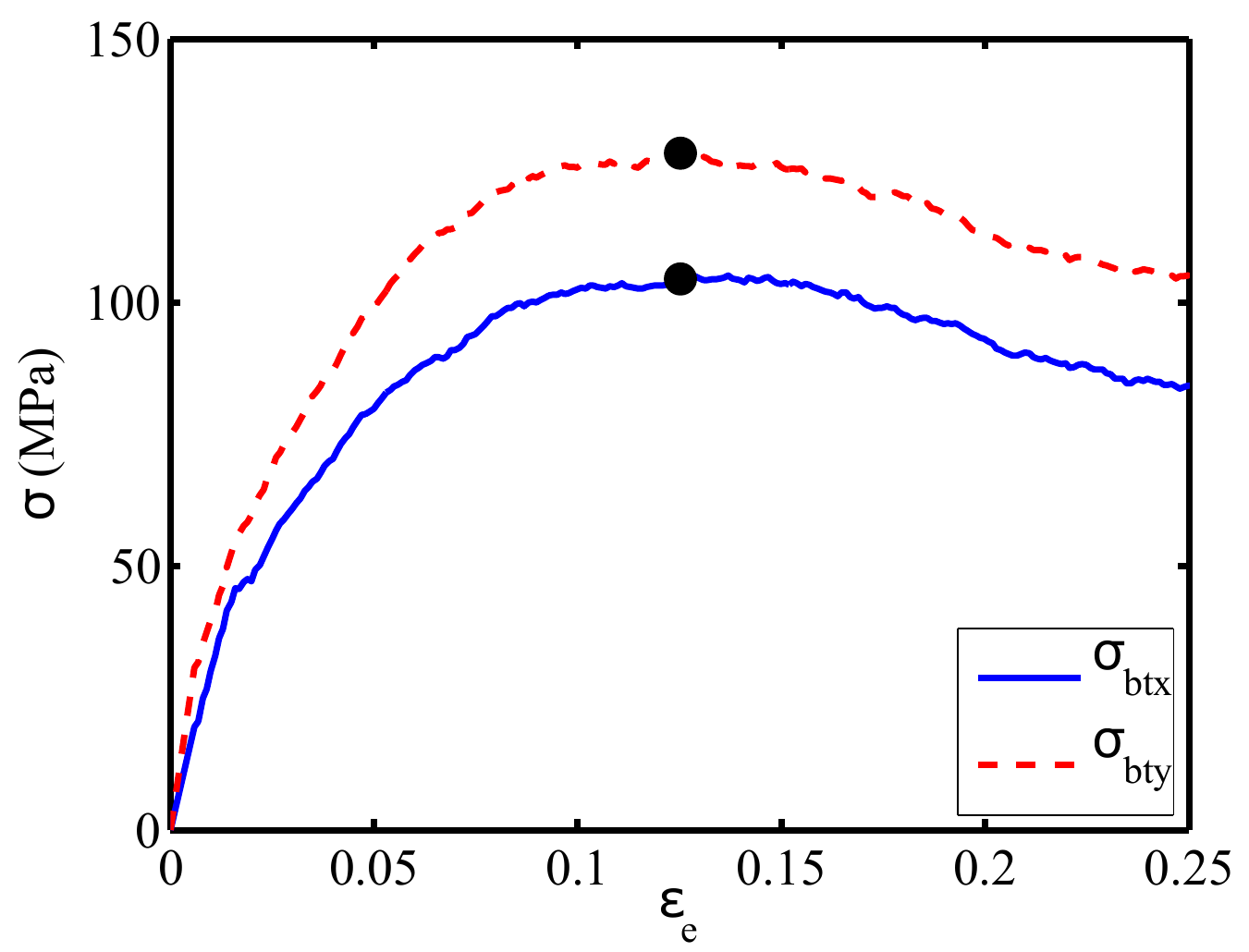}}
	\subfigure[]{\includegraphics[width = 0.45\textwidth]{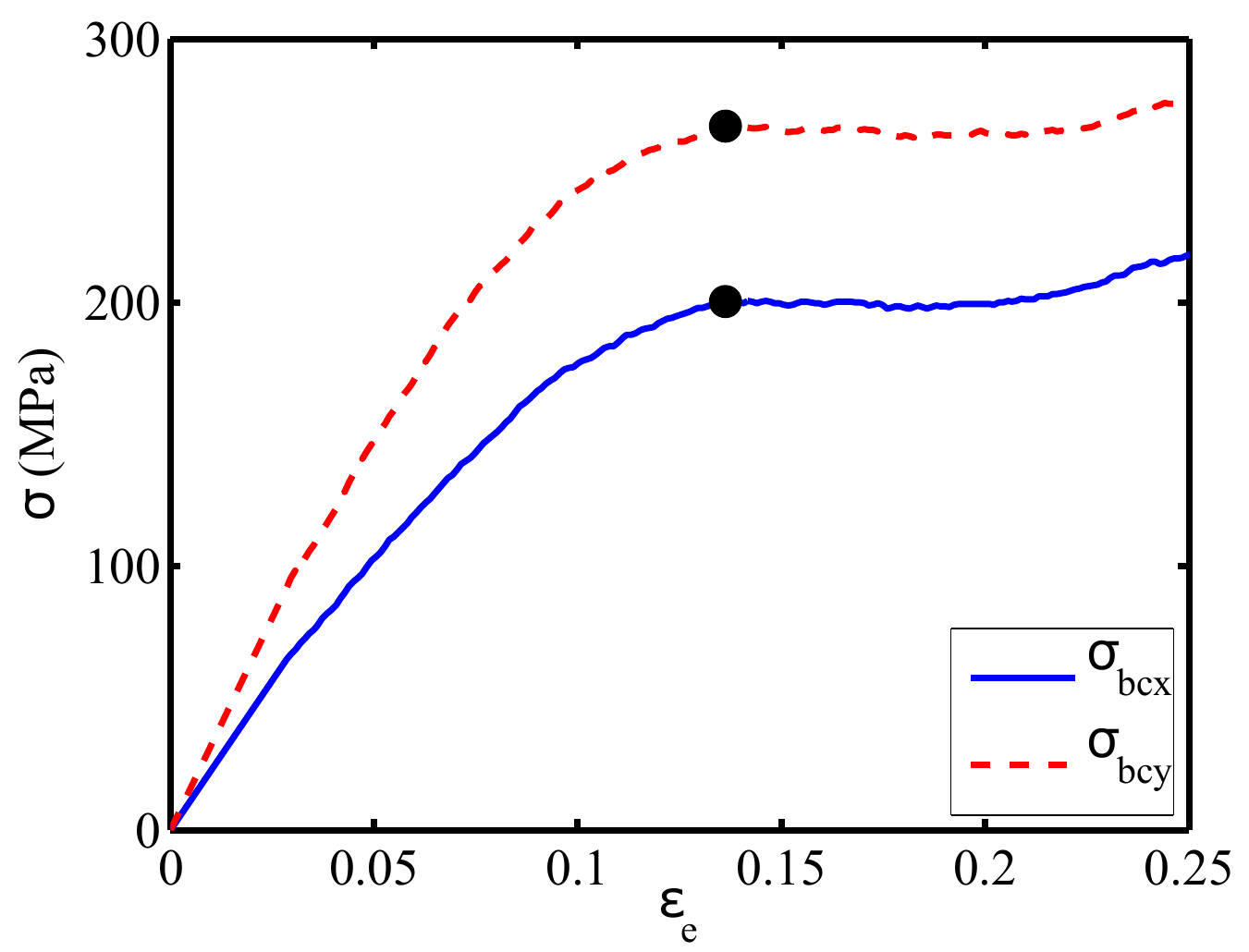}}
	\caption{Principal stress components $\sigma_y ~\text{and} ~\sigma_x$ are obtained from biaxial tension and compression at $T = 300$K with the rates $\dot{\varepsilon}_y = 2 \dot{\varepsilon}_x$. The maximum stresses ($\bullet$) on the curves are indicated as the yield stresses.}
	\label{fig:biax_principal_stresses}
\end{figure}

Figure \ref{fig:biax_principal_stresses} shows the stresses in $x$ and $y$ directions versus the equivalent strain computed by Equation (\ref{eq:equivalent_strain}) and the corresponding yield peaks for biaxial tension and compression. The yield peaks ($\bullet$) occurring at the same equivalent strain are evaluated as maximum stresses.

\subsection{Temperature dependence of elastic moduli}

\begin{figure}[htbp] \centering
	\includegraphics[width = 0.5\textwidth]{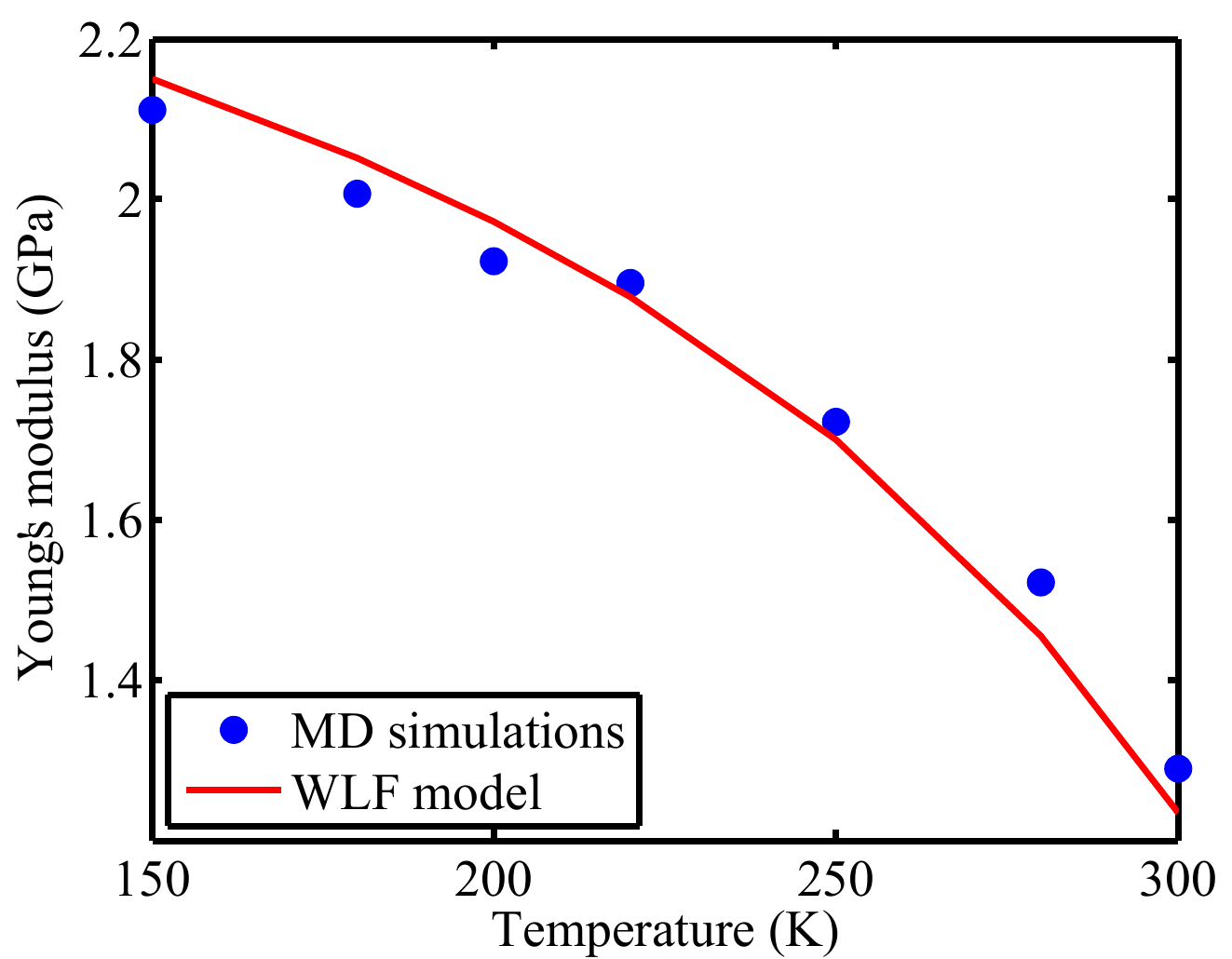}
	\caption{Plot of the tensile Young's modulus versus the temperature. The Young's modulus values are fitted by the solid line formulated in Equation (\ref{eq:wlf_model_modified}).}
	\label{fig:thermo_elastic}
\end{figure}

In order to predict the dependence of the Young's modulus on the temperature the following Williams, Landel and Ferry (WLF) model \cite{Williams:1955} is used:


\begin{equation} \label{eq:wlf_model_modified}
	log a_T = \frac{-C_1 (T - T^{ref} - 140)}{C_2 + (T - T^{ref} - 140)}
\end{equation}

\noindent where $T^{ref}$ is the reference temperature, $C_1$ and $C_2$ are adjustable WLF constants which are calibrated to fit the modulus data. Figure \ref{fig:thermo_elastic} shows that the Young's moduli obtained from MD simulations are well explained by the WLF model.

\subsection{Temperature dependence of yield stresses} \label{subsec:temp_dependence_of_yield_stresses}

Many studies have tried to account for the temperature and strain rate dependence of the yield behavior of polymers. The logarithm law \cite{Eyring:1936}, suggesting the stress-activated jumps of molecular segments results in yielding, is a good candidate to study the dependence of the yield stress on the temperature:

\begin{equation} \label{eq:Eyring_model}
	\sigma = \frac{\Delta H}{V^*} + \frac{R T}{V^*} \ln \frac{2 \dot{\gamma}}{\dot{\gamma}_0}
\end{equation}

\noindent where $\Delta H$ and $V^*$ are the respective activation energy and the activation volume; $\dot{\gamma}$ is the deformation rate, $\dot{\gamma}_0$ is a constant ($\dot{\gamma}_0 \gg \dot{\gamma}$) \cite{Mayr:1998}. The temperature dependent yielding law in Equation (\ref{eq:Eyring_model}) can be approximately substituted by a linear fit (yield stress is considered as a linear function of the temperature) that is used hereafter. Cook et al. \cite{Cook:1998} also reported that the laws used to account for the dependence of yield behavior on the temperature for polymers are mostly linear.

\begin{figure}[htbp] \centering
	\includegraphics[width = 0.5\textwidth]{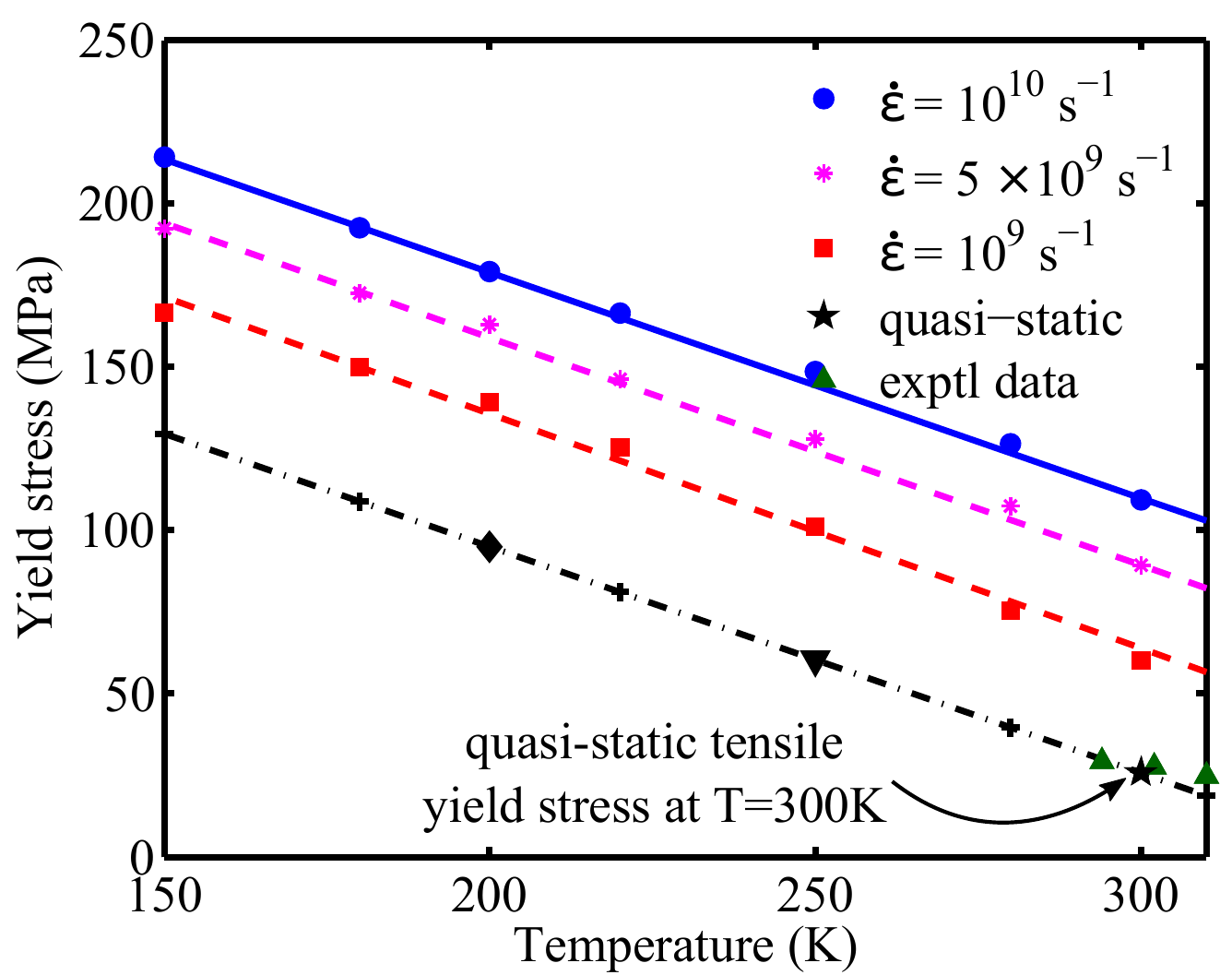}
	\caption{Plot of tensile yield stress obtained from MD simulations as a function of the temperature at different strain rates. The data points are fitted by linear fits according to Equation (\ref{eq:Eyring_model}). The quasi-static tensile yield stress ($\bigstar$) at $T = 300$K is obtained from MD simulations for quasi-static rates in Figure \ref{fig:stress_strain_responses}(c) as presented in Section \ref{subsec:deformations_simulation}. The experimental data are obtained from \cite{Hartmann:1986}.}
	\label{fig:temp_dependent_tens_yield_stress}
\end{figure}

Figure \ref{fig:temp_dependent_tens_yield_stress} shows the curves fitting the yield stress versus temperature for different strain rates are parallel. It means that the slope of the linear fits is nearly rate independent (even at quasi-static conditions). This has also been observed by Rottler et al. \cite{Rottler:2003}. Hence, we are able to predict the yield behavior for different temperatures at quasi-static conditions, if the quasi-static yield stress value at any temperature is provided. Consequently, given the tensile yield stress ($25.92$MPa) at $T = 300$K ($\bigstar$) obtained from MD simulations for quasi-static conditions, see Figure \ref{fig:stress_strain_responses}(c), the temperature dependent yielding law can be constructed at quasi-static strain rates (dash dot black line). A good agreement between this quasi-static linear fit and the experimental yield stresses extracted from \cite{Hartmann:1986} (with strain rate of $2 ~\text{min}^{-1}$) for different temperatures is quite clear. The predicted quasi-static tensile yield stress is related to the tensile yield stress obtained from MD simulations at the temperature of $300$K is expressed by

\begin{equation} \label{eq:tens_yield_stress_scaling_at_T300K}
	\left.\sigma^\text{static}_{t}\right|_{_{T^{ref}}}=\frac{\left.\sigma_{t}\right|_{_{T^{ref}}}}{\gamma} \approx 25.92 ~\text{MPa} \Rightarrow \gamma \approx 0.23
\end{equation}

\begin{figure}[htbp] \centering
	\subfigure[]{\includegraphics[width = 0.485\textwidth]{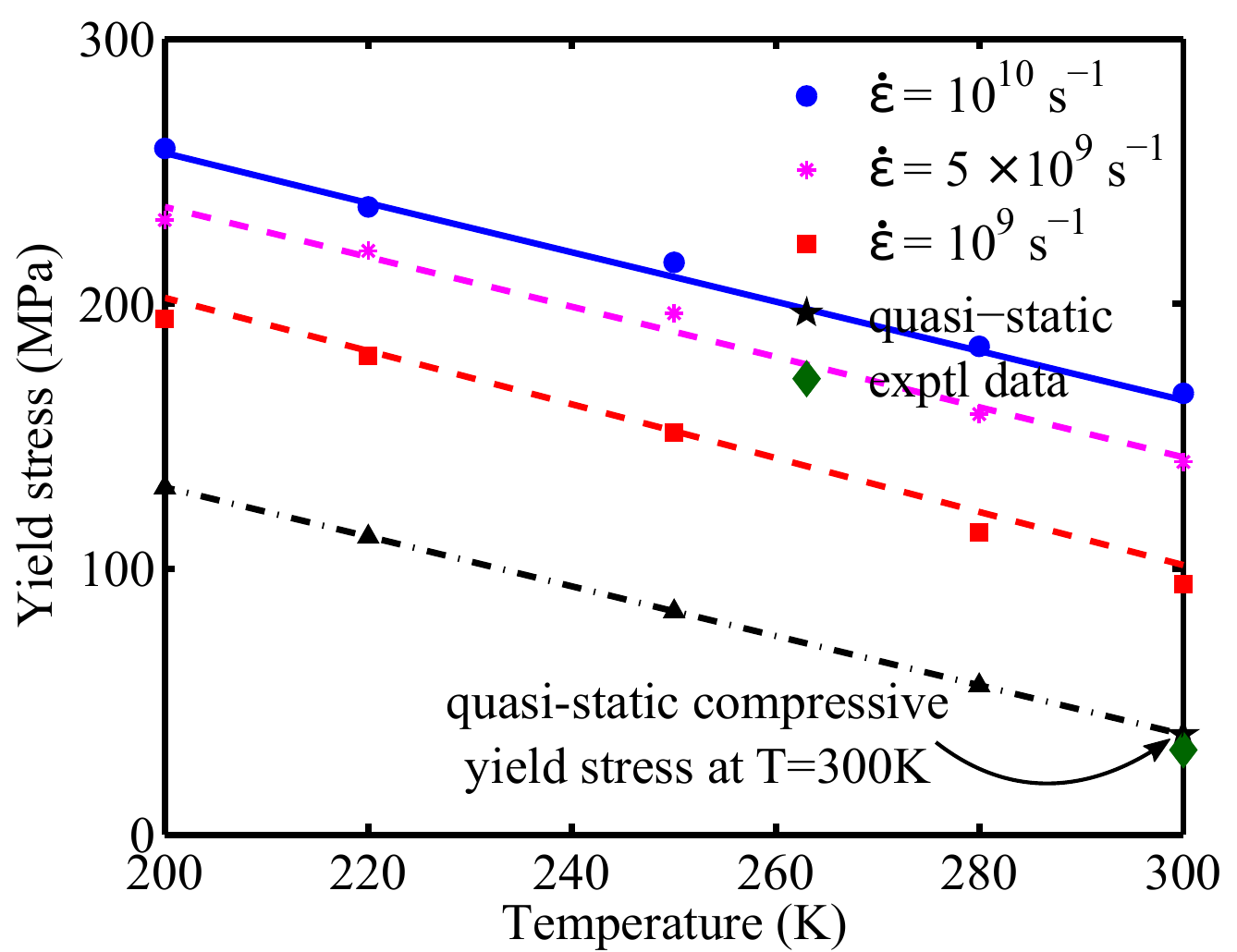}}
	\subfigure[]{\includegraphics[width = 0.485\textwidth]{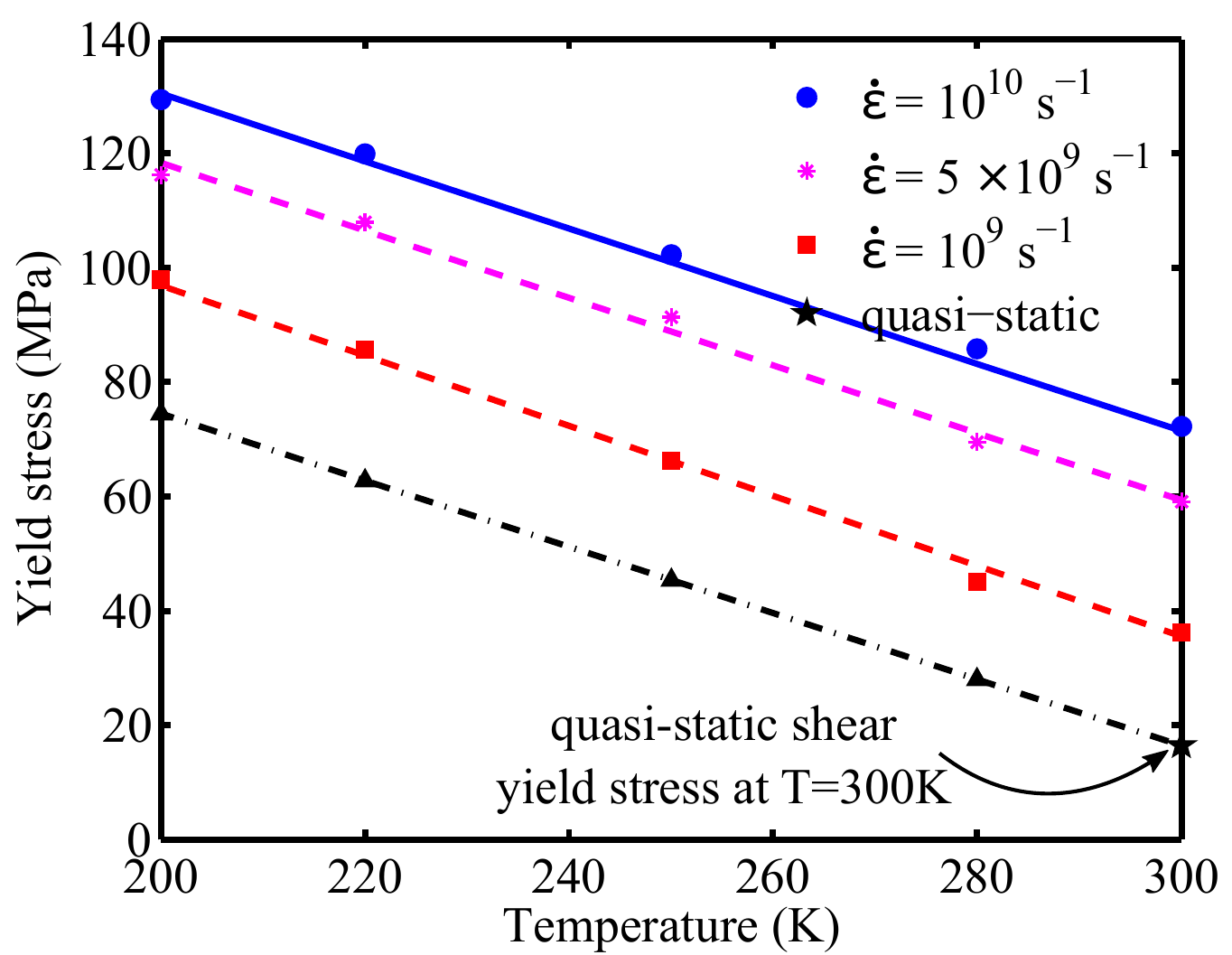}}
	\caption{Temperature dependent (a) compressive yield stress and (b) shear yield stress obtained from MD simulations at different strain rates. The data points are fitted by linear fits. The respective quasi-static compressive ($\bigstar$) and shear yield stresses ($\bigstar$) at $T = 300$K are predicted by scaling the compressive and shear yield stresses obtained from MD simulations by the value given in Equation (\ref{eq:comp_shear_yield_stress_scaling_at_T300K}). The experimentally compressive strength ($\color{OliveGreen}{\blacklozenge}$) obtained from the testing method \emph{Compressive Strength, ASTM D695, $73$F} is $31.7$MPa \cite{QuadrantHDPE,hdpe}.}
	\label{fig:temp_dependent_comp_shear_yield_stress}
\end{figure}

The compressive and shear yield stresses as a function of the temperature for different strain rates are illustrated in Figure \ref{fig:temp_dependent_comp_shear_yield_stress}. The rate dependent the compressive and shear yielding laws obtained from MD simulations also show a parallel behavior. Furthermore, the tensile, compressive and shear laws (fitted lines) at different strain rates approximately change with the same rate suggesting the use of the same scaling factor $\gamma$ to predict the quasi-static compressive and shear yield stresses at the temperature of $300$K. The predicted quasi-static law for compression agrees well to experimental results obtained from ASTM tesing method \cite{QuadrantHDPE,hdpe}, as depicted in Figure \ref{fig:temp_dependent_comp_shear_yield_stress}.

\begin{equation} \label{eq:comp_shear_yield_stress_scaling_at_T300K}
	\left.\sigma^\text{static}_{c}\right|_{_{T^{ref}}}=\frac{\left.\sigma_{c}\right|_{_{T^{ref}}}}{\gamma} \approx -37.6 ~\text{MPa}; \qquad \left.\sigma^\text{static}_{s}\right|_{_{T^{ref}}}=\frac{\left.\sigma_{s}\right|_{_{T^{ref}}}}{\gamma}a \approx 16.34 ~\text{MPa}
\end{equation}

\subsection{Strain rate dependence of the yield stress based on Bayesian approach} \label{subsec:srate_law_based_bayes}

\subsubsection{Bayesian updating}

In this article, Bayesian approach is employed to calibrate the parameters of the plasticity constitutive models from the yield stress data obtained from MD simulations and existing experiments. The advantage of this method is that the naturally uncertain properties of the constitutive parameters existing in the multiscale model for polymers are taken into account. In the Bayes' theorem the random variable $\bm{\theta}$ is expressed by a \emph{prior distribution} $p(\bm{\theta})$. The uncertain parameters being estimated are then directly considered in the model evidence.

\begin{equation} \label{eq:bayes}
	p(\bm{\theta}|\bm{z}) = \frac{p(\bm{z}|\bm{\theta}) p(\bm{\theta})}{p(\bm{z})},
\end{equation}

\noindent with $\bm{\theta},~\bm{z}$ being the vector of model parameters and the vector of observations. For parameter identification purpose, the denominator $p(\bm{z})$ can be ignored and the posterior $p(\bm{\theta}|\bm{z})$is proportionally expressed by a combination of likelihood $p(\bm{z}|\bm{\theta})$ and the prior $p(\bm{\theta})$ as follows:

\begin{equation} \label{eq:posterior}
	\underbrace{p(\bm{\theta}|\bm{z})}_{posterior} \propto \underbrace{p(\bm{z}|\bm{\theta})}_{likelihood} \underbrace{p(\bm{\theta})}_{prior}.
\end{equation}

Subsequently, the parameters yielding the maximum a posterior (MAP) probability of the parameters given the data is identified by:

\begin{equation} \label{eq:MAP}
	\bm{\theta}_{MAP} = \underset{\bm{\theta}}{argmax} p(\bm{z}|\bm{\theta}) p(\bm{\theta}_i)
\end{equation}

\subsubsection{Scaling law constructed based on Bayesian approach}

The strain rate in engineering practice is much lower than the one in MD simulations. Consequently, the respective yield stresses obtained from MD simulations and experiment can differ significantly. Hence, a scaling law is needed to upscale the yield behavior from nanoscale to macroscale. The good agreement between the predicted quasi-static yield stress ($\bullet$) and the experimental results at $T=300$K in Figure \ref{fig:stress_strain_responses}(c) implies that it is possible to rescale the high strain rate involved in MD simulations to macroscopic significant strain rate. Using the above-mentioned Bayesian approach, we can identify parameters of the strain rate dependent law when yield points at different strain rates obtained at molecular and continuum levels are determined. As suggested by earlier researchers \cite{Rottler:2003,Richeton:2006,YangGhosh:2012}, the dependence of the tensile yield stress $\sigma_t$ on the strain rate $\dot{\varepsilon}$ can be described by a logarithm or a power law form. In this article an exponential dependence of the yield stress $\sigma_t$ on the strain rate $\dot{\varepsilon}$ is adopted.


\begin{equation} \label{eq:scaling_law}
\sigma_t = \theta_1 e^{\theta_2 \dot{\varepsilon}} + \theta_3 e^{\theta_4 \dot{\varepsilon}}
\end{equation}

\noindent where $\theta_i, ~i=1,.., 4$ are constitutive parameters calibrated and updated on data points obtained from MD simulations and experimental data.



\begin{figure}[htbp] \centering
	\includegraphics[width = 0.5\textwidth]{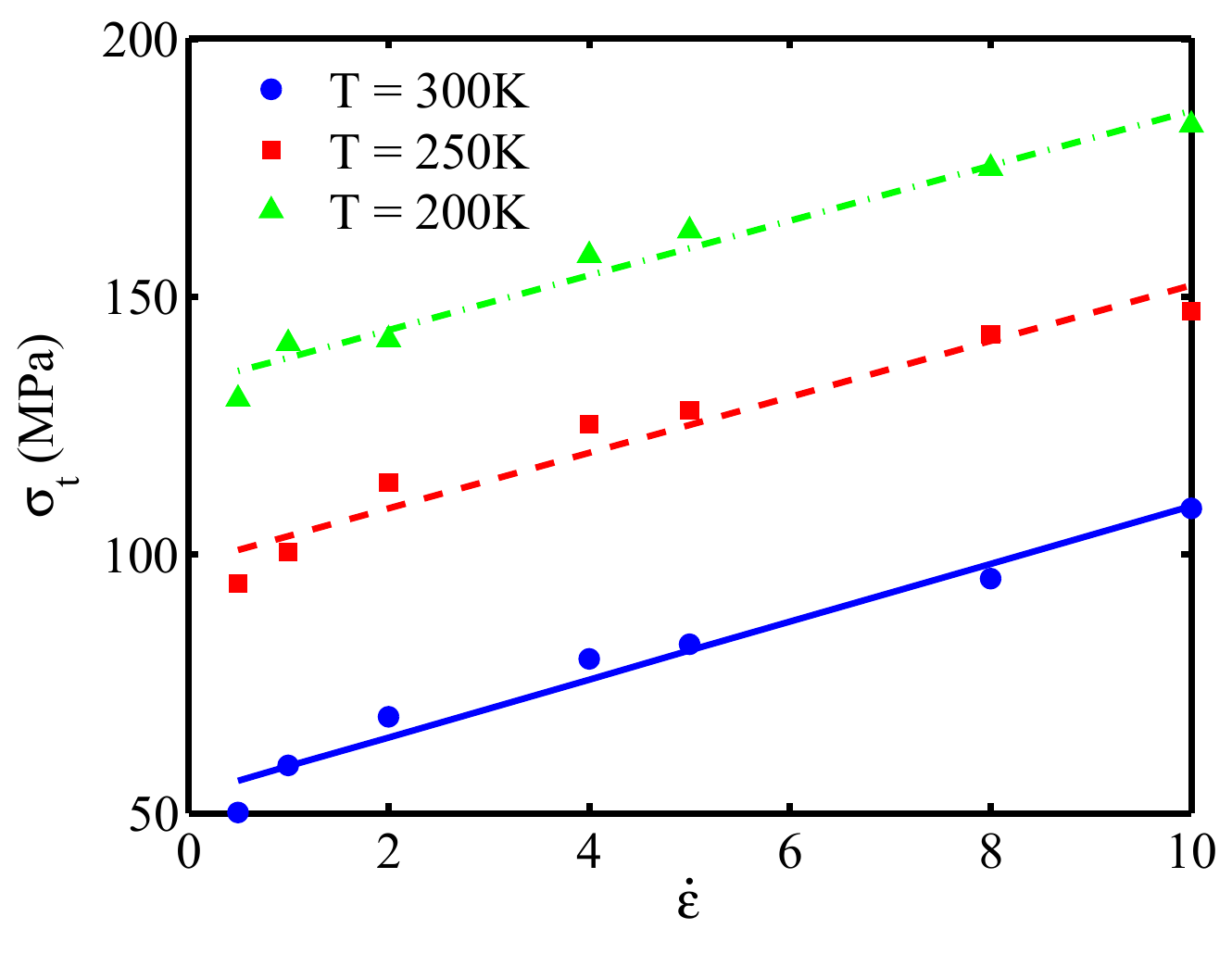}
	\caption{Predicted yield stress as a function of the strain rate for different temperatures at nanoscale model. The data points are fitted by linear functions.}
	\label{fig:srate_dependent_tens_yield_stress}
\end{figure}

\begin{figure}[htbp] \centering
	\includegraphics[width = 0.5\textwidth]{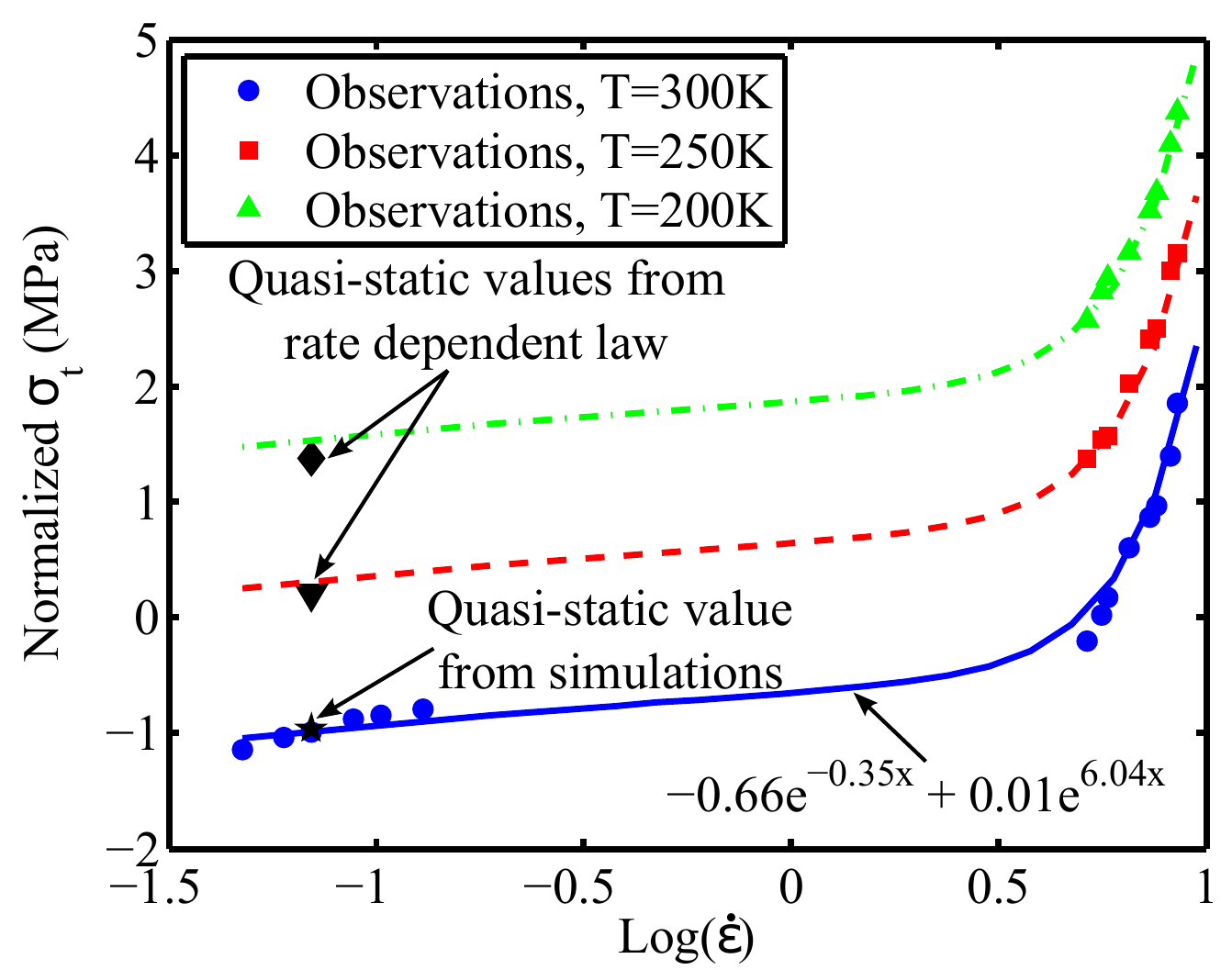}
	\caption{Tensile yield stress versus the logarithm of strain rate ($\log \dot{\varepsilon}$). The blue solid circles represent results obtained from MD simulations and experiments at $T = 300$K. The blue solid line represents the rate dependent yielding law at $T = 300$K obtained by Bayesian approach. The red dashed and green dash dot lines are constructed under the assumption of parallel behavior of the rate dependent yielding law. The black ($\bigstar$) were obtained from MD simulations at quasi-static loading rates, the black asterisk ($\blacktriangledown$) and the black diamond ($\blacklozenge$) are predicted values obtained from the rate dependent yielding law at quasi-static conditions, corresponding to the same symbols ($\blacktriangledown$ and $\blacklozenge$) for $T = 250$K and $T = 200$K in Figure \ref{fig:temp_dependent_tens_yield_stress}, respectively.}
	\label{fig:scaling_law_srate}
\end{figure}

Since the temperature and strain rate are not correlated with respect to (w.r.t.) the yield stress as reported in \cite{NamRabczuk:2012}, a linear transformation of the strain rate dependent yielding law (the numerical fit) for different temperatures is proposed. Richeton et al. \cite{Richeton:2006} also suggested that the fitted law can be linearly transformed in vertical and horizontal directions when considering the effect of temperature and strain rate, respectively. Rate dependence of the yielding law is studied for different temperatures at the nanoscale model. Interestingly, as can be seen in Figure \ref{fig:srate_dependent_tens_yield_stress},  fitted curves to the yield points at different temperatures are parallel. This supports the assumption that the rate dependent yielding law has also parallel behavior even at low strain rates. It means that under this assumption the rate dependence of yield stress for different temperatures can be predicted on a large ranges of strain rate (from low rate in practical application to high rate involved in MD simulations). As shown in Figure \ref{fig:scaling_law_srate}, predictions for the rate dependence of the yield stress at $T=250$K and $T=200$K are possible. The functional form in Equation (\ref{eq:scaling_law}) provides a good fit to the data. Good agreement of the predicted law with quasi-static yield stresses at $T=250$K and $T=200$K is observed. Hence, the model parameters can be identified in the case of limited experimental data from a Bayesian perspective. Note that the predicted quasi-static tensile yield stresses at $T = 250$K ($\blacktriangledown$) and $T = 200$K ($\blacklozenge$) correspond to the values illustrated by the same symbols ($\blacktriangledown$ and $\blacklozenge$) on the quasi-static fitted curve (black dash dot curve) in Figure \ref{fig:temp_dependent_tens_yield_stress}, respectively.

\section{Macroscopic Continum model} \label{sec:macroscale}

\subsection{Definition of yield surface}

\begin{figure}[htbp] \centering
	\includegraphics[width = 0.5\textwidth]{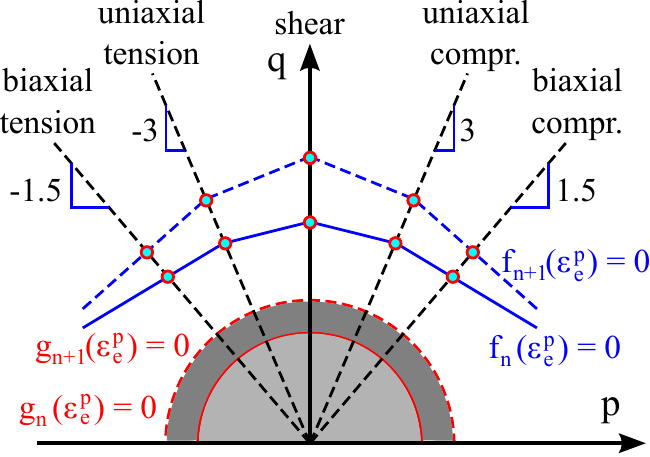}
	\caption{Plot of piecewise linear yield surface (PLYS) in invariant ($p,q$) plane.}
	\label{fig:PLYS}
\end{figure}

The yield surface requires five constitutive properties: the uni- and biaxial tensile, compressive, and shear yield strength obtained from MD simulations. It is then constructed up to four Drucker-Prager-cones as suggested by Vogler \emph{et al.} \cite{Vogler:2007}:

\begin{equation} \label{eq:multisurf_yieldfuncs}
	f (p, q, \varepsilon^p_e ) = q - \beta (\varepsilon^p_e) p -c (\varepsilon^p_e)
\end{equation}

\noindent where $q = \sqrt{ 3 J_2 } = \sqrt{ \frac{3}{2} \mathbf{s:s} }$ is the von-Mises equivalent stress; $p = -\frac{1}{3} I_1$ is the hydrostatic pressure, with ${I_1} = tr(\boldsymbol{\sigma})$ being the first stress invariant, and ${J_2}=\frac{1}{2} {\boldsymbol{s}}:{\boldsymbol{s}}$ being the second invariant of the deviatoric stress tensor $\boldsymbol{s}$ while $\varepsilon^p_e$ is the equivalent plastic strain. The parameter $\beta$ can be expressed in terms of the equivalent plastic strain \cite{Vogler:2007} as

\begin{equation} \label{eq:multsurf_params}
\begin{aligned}
\beta (\varepsilon^p_e) & = 3 \frac{\sigma_t - \sigma_{bt}}{2 \sigma_{bt} - \sigma_t}, \quad c (\varepsilon^p_e) = \sigma_t + \beta (\varepsilon^p_e) \frac{\sigma_t}{3} \quad \text{for} \quad p < - \frac{q}{3} \\
\beta (\varepsilon^p_e) & = 3 \frac{\sqrt{3} \sigma_s - \sigma_t}{\sigma_t}, \quad c (\varepsilon^p_e) = \sqrt{3} \sigma_s  \quad \text{for} \quad - \frac{q}{3} \leq p < 0 \\
\beta (\varepsilon^p_e) & = 3 \frac{\sigma_c - \sqrt{3} \sigma_s}{\sigma_c}, \quad c (\varepsilon^p_e) = \sqrt{3} \sigma_s  \quad \text{for} \quad 0 \leq p < \frac{q}{3} \\
\beta (\varepsilon^p_e) & = 3 \frac{\sigma_{bc} - \sigma_c}{2 \sigma_{bc} - \sigma_c}, \quad c (\varepsilon^p_e) = \sigma_c - \beta (\varepsilon^p_e) \frac{\sigma_c}{3} \quad \text{for} \quad p \geq \frac{q}{3}
\end{aligned}
\end{equation}

\noindent where the parameters $\beta (\varepsilon^p_e)$ is extracted from the hardening uniaxial tensile ($\sigma_t$), uniaxial compressive ($\sigma_c$), shear ($\sigma_s$), biaxial tensile ($\sigma_{bt}$), and biaxial compressive ($\sigma_{bc}$) curves which are obtained from MD simulations for corresponding stress states. The piecewise linear yield surface (PLYS) is illustrated in Figure \ref{fig:PLYS}.

In the proposed model, a non-associated flow rule is used to ensure the consistency with tensile test, leading to the plastic potential suggested by \cite{NamLiu:2015}:



\begin{equation} \label{eq:plas_pot}
	g = q^2 + \alpha p^2
\end{equation}

\noindent where $\alpha$ is the flow parameter accounting for the change in material volume at yielding:

\begin{equation} \label{eq:flow_param}
	\alpha = \frac{9}{2} \frac{1-2 \nu_p}{1+ \nu_p}
\end{equation}

\noindent where $\nu_p$ denotes the plastic Poisson's ratio obtained from the MD simulations under uniaxial tension and

\begin{equation}\label{eq:matrix_nup}
	\Delta \varepsilon^p_{22} = \Delta \varepsilon^p_{33} =-\nu_p\Delta \varepsilon^p_{11}
\end{equation}

The increment of plastic deformation is given by:

\begin{equation} \label{eq:plas_deform}
	\Delta \bm{\varepsilon}^p = \Delta \lambda \frac{\partial g}{\partial \boldsymbol\sigma}
\end{equation}

\noindent where $\Delta \lambda$ is the plastic multiplier, commonly updated via the return mapping algorithm under the Kuhn-Tucker consistency conditions. A more efficient approach based on Chen Mangasarian replacement functions which avoids a return mapping has been proposed by \cite{Areias:2012,Areias:2015}; $\frac{\partial g}{\partial \boldsymbol\sigma}$ represents the direction of plastic flow with $g$ being the plastic potential given in Equation (\ref{eq:plas_pot}). The equivalent plastic strain is given by \cite{Melro:2013a}:

\begin{equation} \label{eq:eq_plastic_strain}
\varepsilon^p_e= \sqrt{k \varepsilon^p \colon \varepsilon^p}
\end{equation}

\noindent with $k=\frac{1}{1+2\nu_p^2}$.

\subsection{Thermo-plastic hardening} \label{subsec:hardening_law}

The constitutive model is defined by uniaxial tension and compression, biaxial tension and compression and shear yield strengths. Thus, the hardening will be formulated to update these yield strengths. Commonly to other plasticity models, the hardening formulation depend on the equivalent plastic strain as follows:

\begin{equation} \label{eq:hardening_laws}
	\sigma_t = \sigma_t \left( \varepsilon^p_e \right), \quad \sigma_c = \sigma_c \left( \varepsilon^p_e \right), \quad \sigma_s = \sigma_s \left( \varepsilon^p_e \right), \quad \sigma_{bt} = \sigma_{bt} \left( \varepsilon^p_e \right), \quad \sigma_{bc} = \sigma_{bc} \left( \varepsilon^p_e \right)
\end{equation}

We can directly extract stress and strain values from the uni- and biaxial tension and compression and shear from MD simulations. Then, the hardening laws were inserted into the material model in terms of table of values. Note that input data are presented in terms of plastic strain by decomposing the total strain increment by the elastic component as: $\Delta \varepsilon^p = \Delta \varepsilon - \Delta \varepsilon^{el}$ \cite{NamLiu:2015}. In each iteration the table lookups will provide the plastic strains $(\varepsilon^p_e)$ and corresponding yield stresses $(\sigma_y)$ as inputs. Subsequently, the tangents $\left( \frac{\partial \sigma_y}{\partial \varepsilon^p_e} \right)$ with respect to the plastic strain will be computed. These stress-plastic strain curves are then scaled to determine the hardening laws. In order to study the temperature dependent yield strength, the linear law fitted on data obtained from MD simulations is employed to scale the yield stress w.r.t. the hardening curve at the reference temperature as follows:

\begin{equation} \label{eq:temp_scaling}
	\sigma_n=\sigma^{ref}_n + \beta_n  \left(T-T^{ref}\right)
\end{equation}

\noindent with $\sigma_n$ and $\sigma^{ref}_n$ being the predicted yield stresses at the desired $T$  and reference $T^{ref}$ temperatures, respectively. The material constant $\beta_n$ is selected so that the yield stresses at the temperature $T$ are scaled back to the stresses' value at the reference temperature $T^{ref}$. An overview illustrating the algorithm that is applied to implement the PLYS constitutive model is shown in Table \ref{tab:algorithm}.

\bgroup
\def\arraystretch{1}
\begin{table}[!ht]
\begin{threeparttable}
	\caption{Multisurface constitutive model algorithm overview for PE.}
	\label{tab:algorithm}
	\begin{tabular}{|ll|}
		\hline
		(1)	& Compute trial stress, \\
		
		&$\quad \boldsymbol\sigma_{n+1}^{tr} = \boldsymbol\sigma_{n} +\boldsymbol D^e : \Delta  \boldsymbol \varepsilon $ \\
		
		& representing the stress in terms of the von-Mises equivalent $q_{n+1}^{tr}$ and hydrostatic $p_{n+1}^{tr}$ stresses: \\
		&$q_{n+1}^{tr} = \sqrt{\frac{3}{2} \boldsymbol s_{n+1}^{tr}:\boldsymbol s_{n+1}^{tr}}$, $p_{n+1}^{tr} = p_{n} + K \Delta  \varepsilon_v $ with $\quad \boldsymbol s_{n+1}^{tr} = \boldsymbol s_{n} + 2G \Delta   \boldsymbol\varepsilon_d $ \\[5pt]
		
		(2)	& Consider the temperature dependent elastic and yield behavior: \\
		& $\quad$ The temperature ($T$) dependence of the Young's modulus is explained by Equation (\ref{eq:wlf_model_modified}): \\
		& $\qquad E(T) = E^{ref} \left( log a_T \right) $ \\
		& $\quad$ The yield stresses and hardening laws dependent on the temperature ($T$) \\
		& $\quad$ is described by Equation (\ref{eq:temp_scaling}): \\
		& $\qquad \sigma_{n}=\sigma_{n}^{ref} + \beta_n  \left(T-T^{ref}\right)$\\[5pt]
		(3)	& Check yield criterion given by Equation (\ref{eq:multisurf_yieldfuncs}): \\
		& IF $f \left( p,q,\varepsilon^p_e \right)\leq 0$ THEN \\
		& $\quad \boldsymbol\sigma_{n+1}  = \boldsymbol\sigma_{n+1}^{tr} $, $ q_{n+1}  = q_{n+1}^{tr} $, $ p_{n+1}  = p_{n+1}^{tr} $ and EXIT \\[5pt]
		
		&ELSE \\
		& $\quad$ Perform return mapping algorithm to obtain plastic multiplier $\Delta \lambda$. \\
		&ENDIF \\[5pt]
		
		(4) 	&Update stress tensor \\
		& $\quad q_{n+1} = \sqrt{ \frac{3 J^{tr}_2}{(1+6G \Delta \lambda)^2}}$, $ \boldsymbol s_{n+1} = \frac{\boldsymbol s_{n+1}^{tr}}{1+6G \Delta \lambda} $,  $ p_{n+1} = \frac{p_{n+1}^{tr}}{1+2K \alpha  \Delta \lambda} $ \\
		& $\quad \boldsymbol\sigma_{n+1}  = \boldsymbol\sigma_{n+1}^{tr} - 6G\Delta \lambda \boldsymbol s_{n+1} - \frac{2}{3} K \alpha \Delta \lambda p_{n+1} \boldsymbol I $ \\ [5pt]
		(5)	&EXIT \\
		\hline
	\end{tabular}
	\begin{tablenotes}
    \item[\textdagger] The superscript $^{ref}$ is used to infer the quantities computed at the reference temperature.
    \item[\textdagger] $\Delta \varepsilon_v$ and $\Delta \varepsilon_d$ are the volumetric and deviatoric plastic strain increments, respectively.
    \end{tablenotes}
\end{threeparttable}

\end{table}
\egroup

\subsection{Yield surface at different temperatures} \label{subsec:yieldsurf_at_difftemp}

\begin{figure}[htbp] \centering
	\includegraphics[width = 0.5\textwidth]{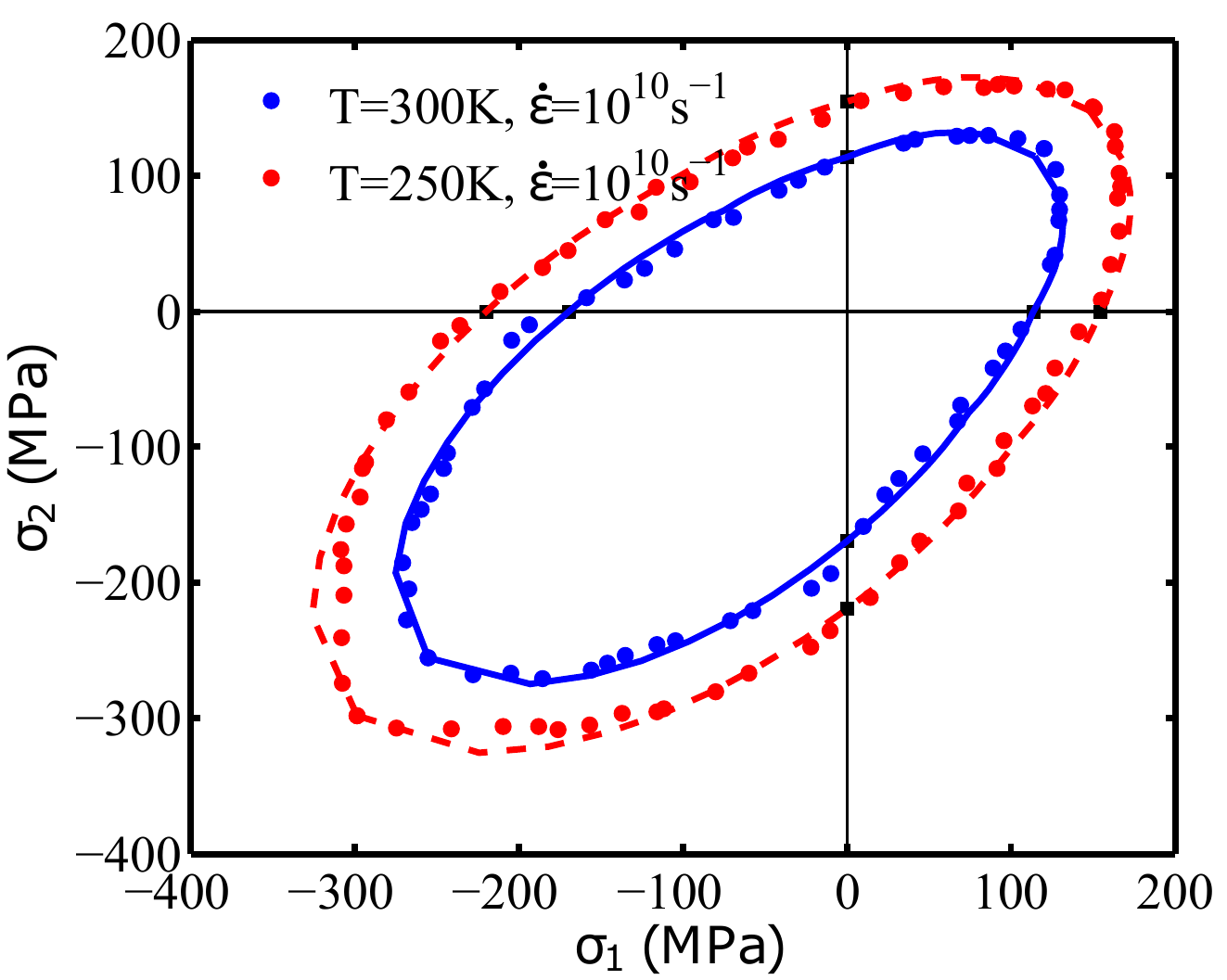}
	\caption{Yield points for uni- and biaxial stress states and the fitted PLYS for different temperatures.}
	\label{fig:temp_dependt_multsurf}
\end{figure}

We perform multiaxial deformations (uni- and biaxial and shear loads) to obtain yield points at two different temperatures and the equivalent strain rate in Equation (\ref{eq:equivalent_strain}) is set as $\dot{\varepsilon}_e = 1 \times 10^{10} ~s^{-1}$. The PLYS characterized by Equation (\ref{eq:multisurf_yieldfuncs}) was adopted to fit yield points data in four Drucker-Prager cones as mentioned in Equation (\ref{eq:multsurf_params}). As can be seen in Figure \ref{fig:temp_dependt_multsurf}, the yield points are well described by the PLYS criterion.

\subsection{Yield surface at different strain rates} \label{subsec:yieldsurf_at_diffsrates}

\begin{figure}[htbp] \centering
	\subfigure[]{\includegraphics[width = 0.45\textwidth]{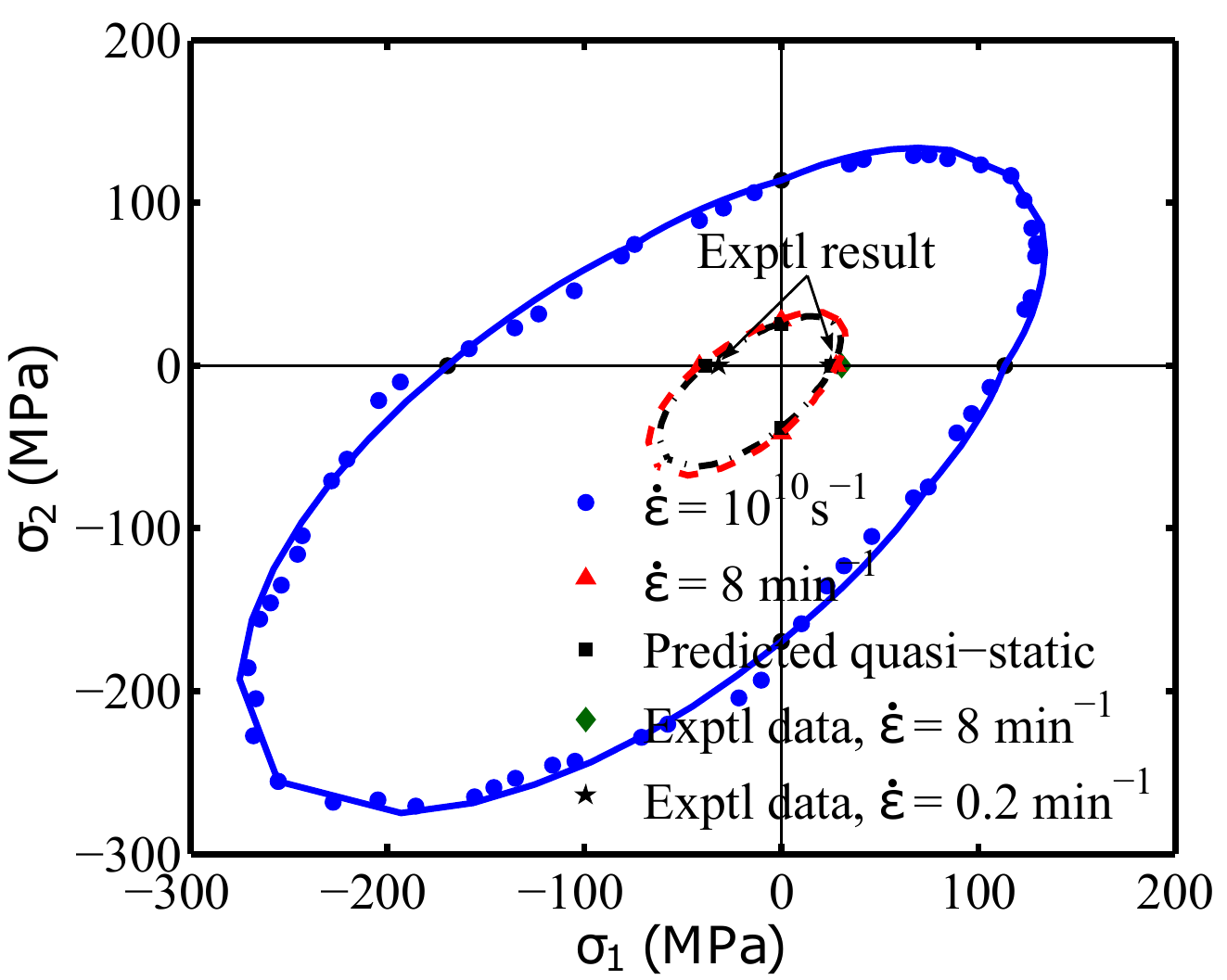}}
	\subfigure[]{\includegraphics[width = 0.45\textwidth]{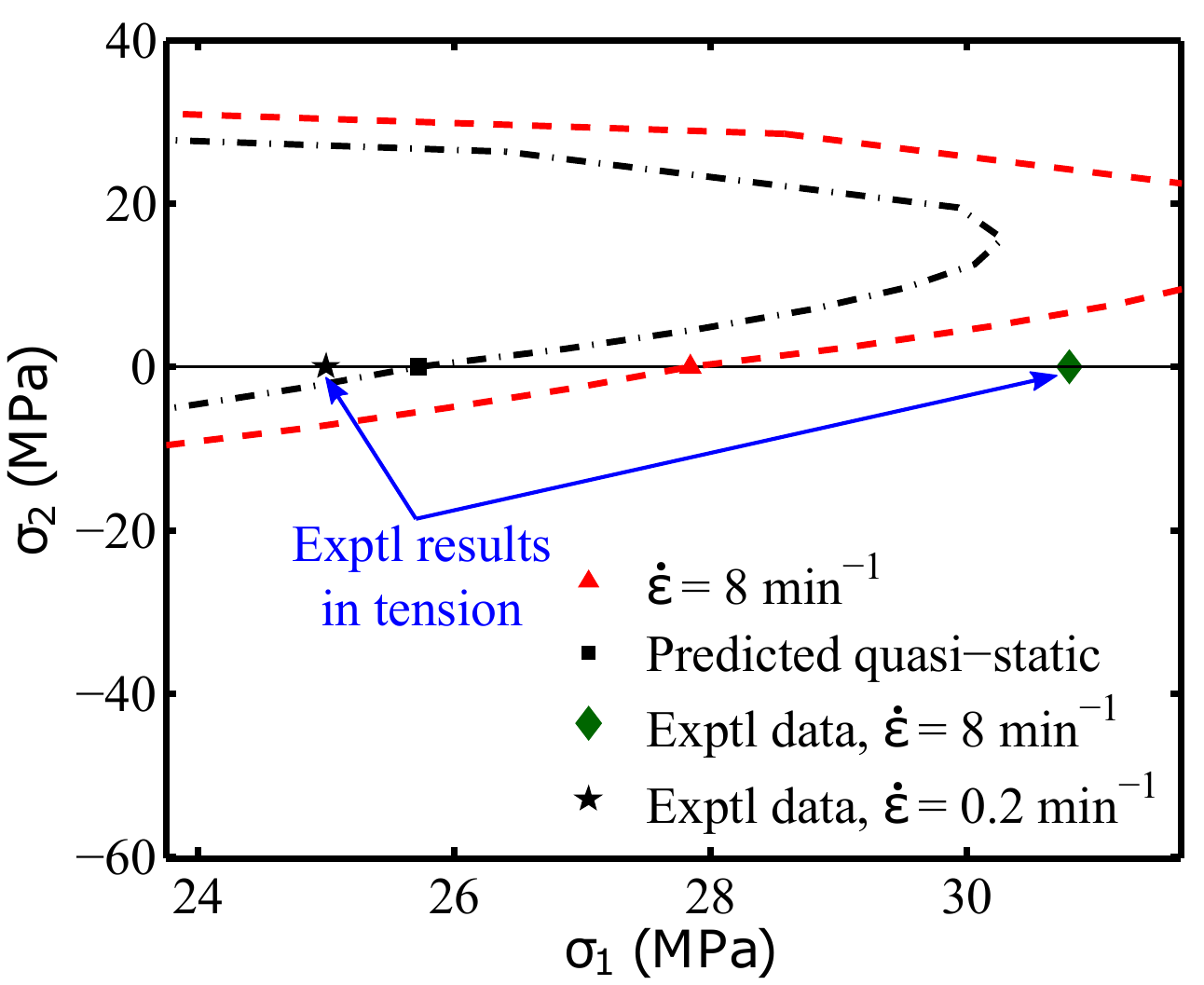}}
	\subfigure[]{\includegraphics[width = 0.45\textwidth]{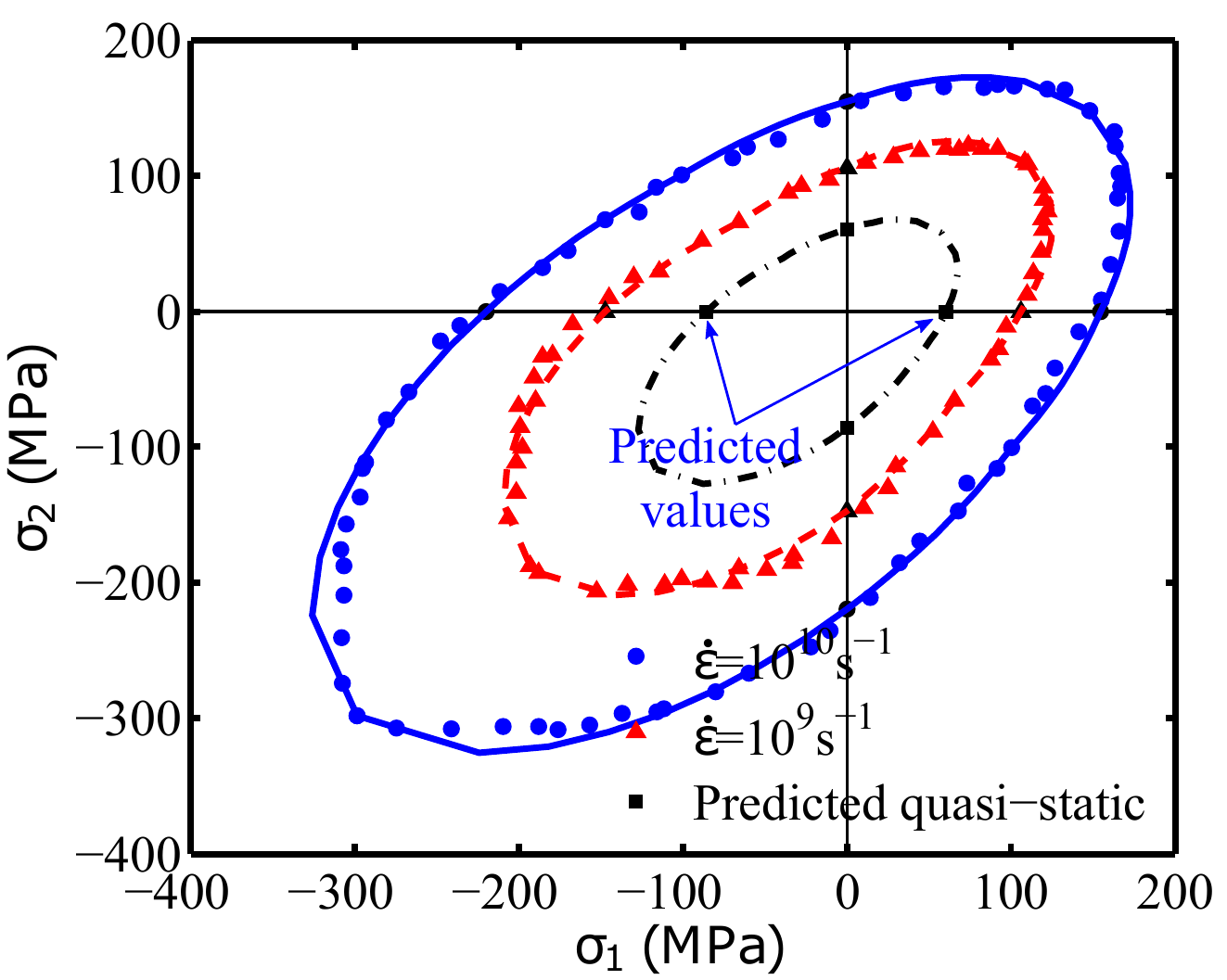}}
	\caption{Yield points for uni- and biaxial stress states and the fitted PLYS (a) for $T = 300$K; (b) the predicted quasi-static tensile yield stresses in comparison with experimental data; (c) for $T = 250$K at different strain rates.} 
	\label{fig:srate_dependt_multsurf}
\end{figure}

Based on the scaling law proposed in Section \ref{subsec:srate_law_based_bayes}, given any known (predicted) yield stress of a specific load case, the entire multisurface yield functions can be isotropically scaled to quasi-static rates by assuming the scaling value is similar for general deformations. For example, a prediction for the entire quasi-static multisurface yield functions at $T = 300$K is obtained in Figure \ref{fig:srate_dependt_multsurf}(a). The uniaxially quasi-static tensile ($25.92$MPa) and compressive ($37.6$MPa) yield stresses are validated with experimental results. As observed in Figure \ref{fig:srate_dependt_multsurf}(a) and Table \ref{tab:validate_quasistatic_results}, good agreement between numerical results and experimental results is observed. Furthermore, the entire yield surface at any desired strain rate can also be predicted based on the law shown in Figure \ref{fig:scaling_law_srate} using Equations (\ref{eq:tens_yield_stress_scaling_at_T300K} + \ref{eq:comp_shear_yield_stress_scaling_at_T300K}). Also, the entire yield surfaces at $T = 250$K for different strain rates are obtained from MD simulations and the one at quasi-static rates is predicted using the same scaling law as illustrated in Figure \ref{fig:srate_dependt_multsurf}(b).

\bgroup
\def\arraystretch{1}
\begin{table}[!ht] 
	\caption{Validation of the predicted tensile and compressive yield stresses with experimental result at different strain rates.} 
	\label{tab:validate_quasistatic_results}
	\begin{center}
		\begin{tabular}{l c c c }
			Deformation & Strain rate	& Quasi-static simulations	& Experimental results \\
			\hline
			\multirow{2}{*}{Tension}	& 0.2 $min^{-1}$	& 25.92 (MPa)& 25.0 (MPa) \cite{Hartmann:1986} \\
			& 8.0 $min^{-1}$	& 28.84 (MPa) & 30.8 (MPa) \cite{Hartmann:1986} \\
			\hline
			Compression	& ~	& - 37.61 (MPa) & -31.72 (MPa) \cite{QuadrantHDPE} \\
			\hline
		\end{tabular}
	\end{center}
\end{table}
\egroup

The above-described elasto-plastic model is used to predict the thermoplastic behavior at (1) nanoscale and (2) the strain rate, which is rescaled from molecular to continuum levels through the constitutive law.

\bgroup
\def\arraystretch{2}
\begin{table}[!ht] 
	\caption{Constitutive properties for the PE model obtained from MD simulations at the room temperature.} 
	\label{tab:PE_constitutive_props}
	\begin{center}
		\begin{tabular}{c c c c c c c c}
			$E$	& $\nu$	& $\nu_p$	& $\sigma^{\text{static}^{ref}}_{t}$	& $\sigma^{\text{static}^{ref}}_{c}$	& $\sigma^{\text{static}^{ref}}_{s}$	& $\beta_T$	& $\beta_C$ \\
			\hline
			1.32 GPa	& 0.32	& 0.32	& 25.92 MPa	& -37.61 MPa	& 16.34 MPa	& 0.55	& 0.93 \\
			\hline
			$\beta_S$	& $C_1$	& $C_2$	& $\theta_1$	& $\theta_2$ & $\theta_3$	& $\theta_4$ \\
			\hline
			0.58		& 0.44	& 82.5	& -0.66	& -0.35	& 0.01	& 6.04 \\
		\end{tabular}
	\end{center}
\end{table}
\egroup

\begin{figure}[htbp] \centering
	\subfigure[]{\includegraphics[width = 0.45\textwidth]{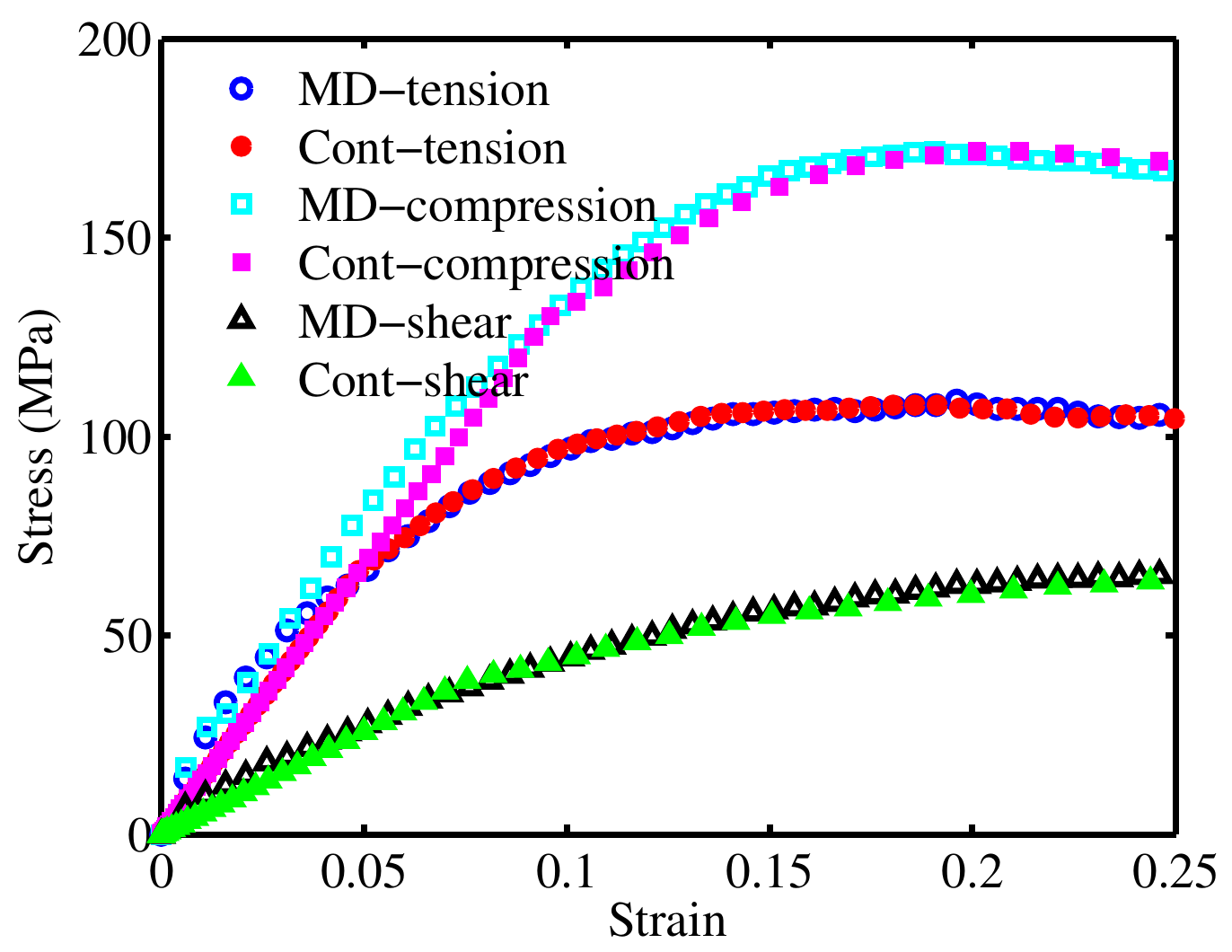}}
	\subfigure[]{\includegraphics[width = 0.45\textwidth]{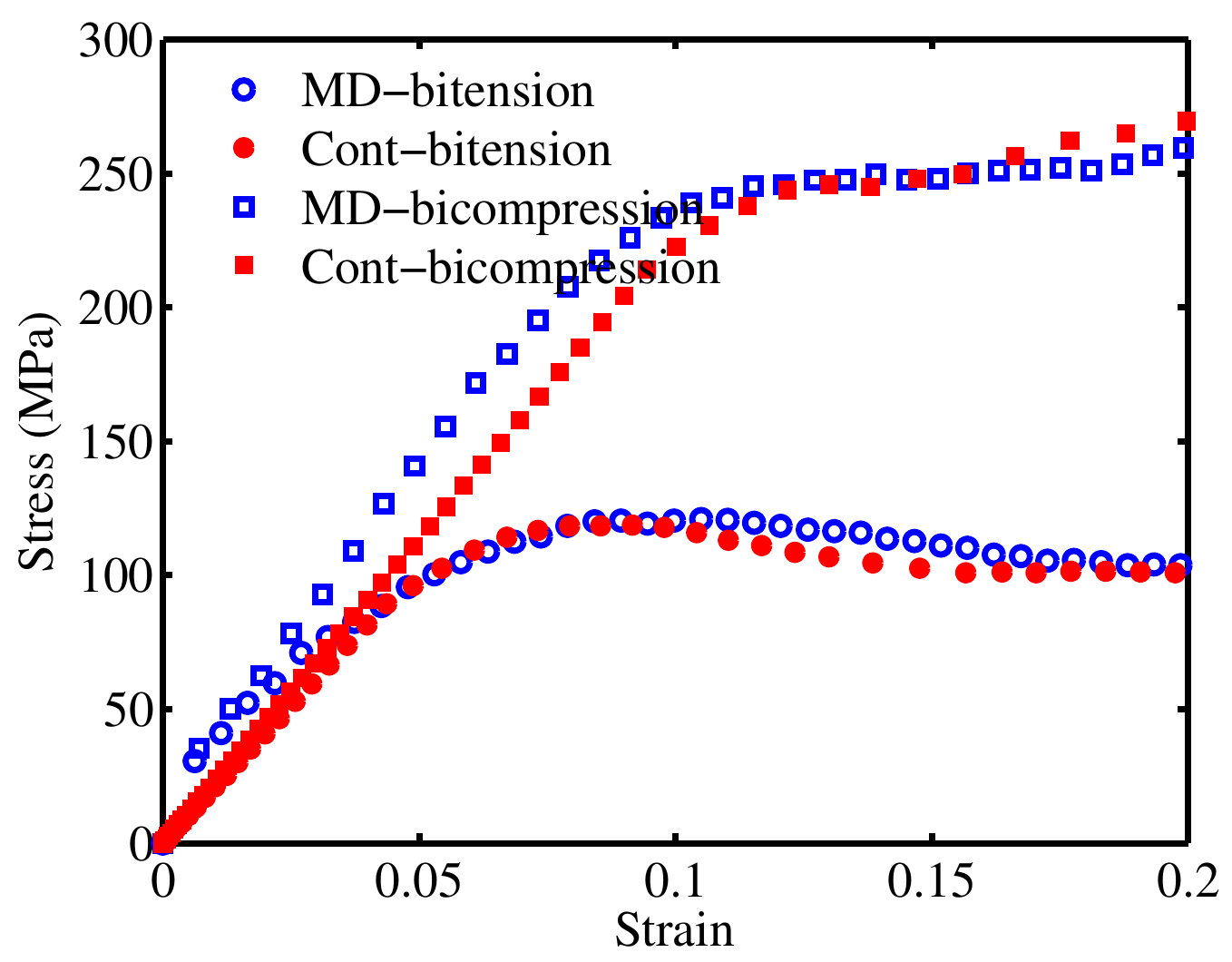}}
	\subfigure[]{\includegraphics[width = 0.45\textwidth]{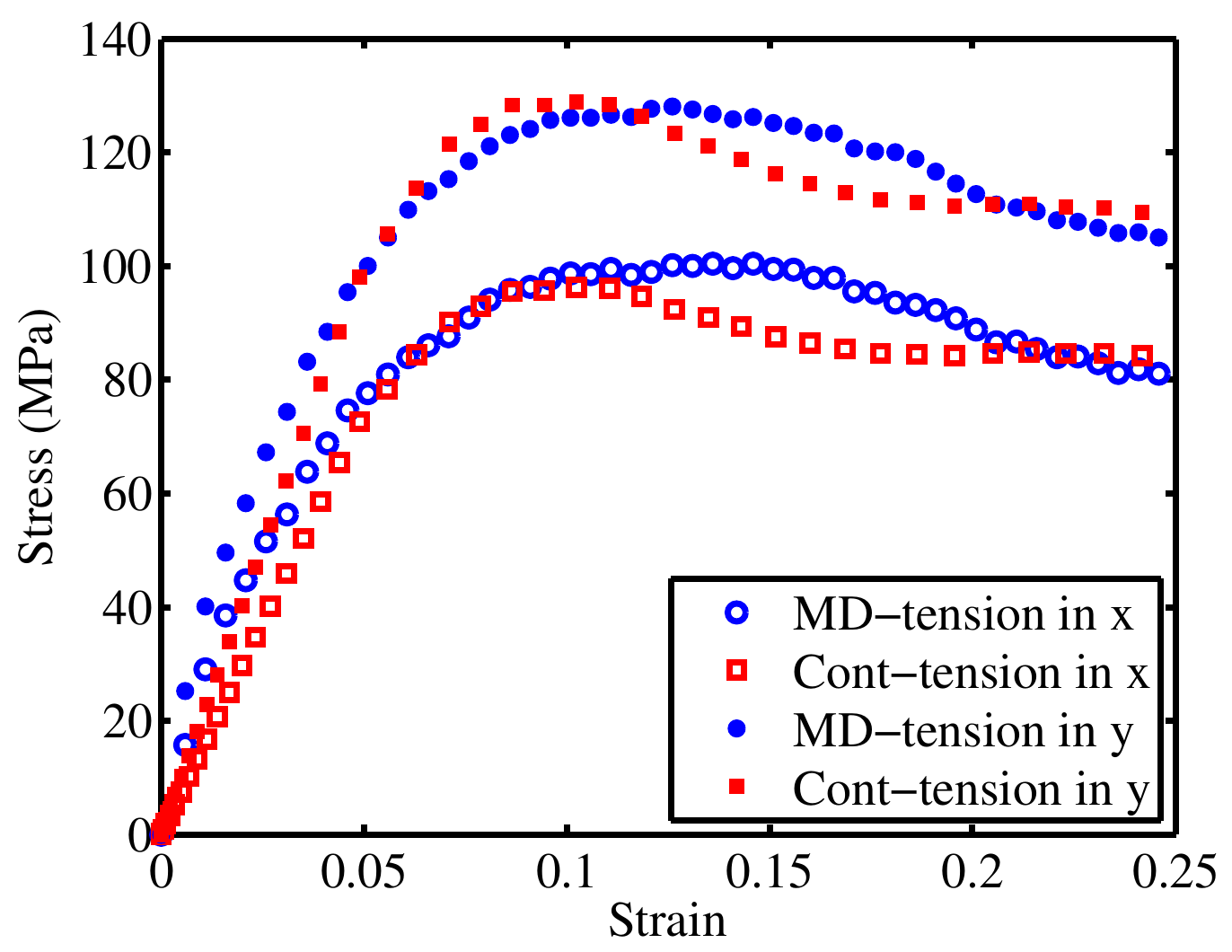}}
	\subfigure[]{\includegraphics[width = 0.45\textwidth]{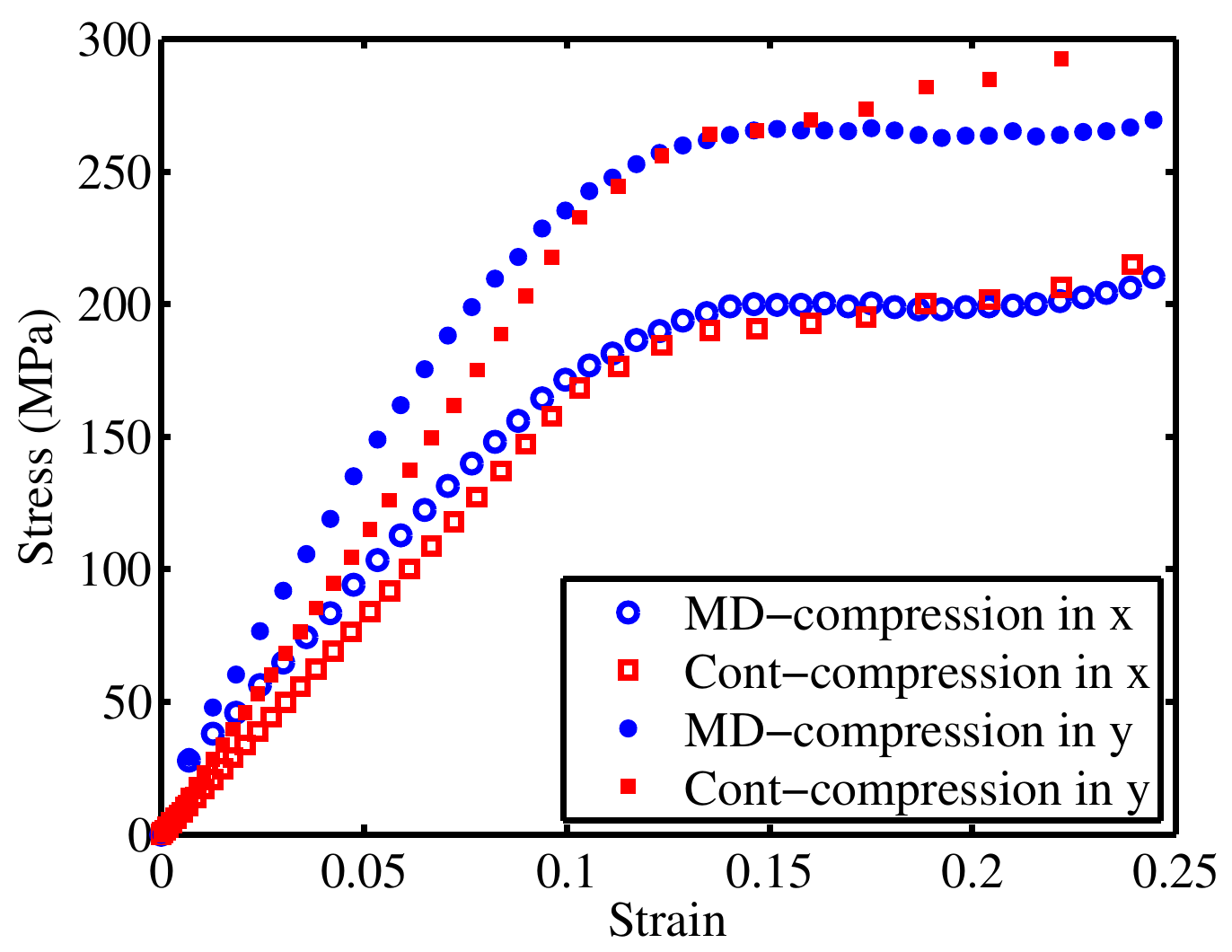}}
	\caption{Comparison of the stress-strain responses predicted by the continuum model and MD simulations (a) tension, compression and shear, (b) biaxial-tension and compression with equal rates applied in $x$ and $y$ directions, (c) biaxial-tension with rates applied in $x$ and $y$ directions $\dot{\varepsilon}_y = 2 \dot{\varepsilon}_x$, (d) biaxial-compression with rates applied in $x$ and $y$ directions $\dot{\varepsilon}_y = 2 \dot{\varepsilon}_x$.}
	\label{fig:verify_stress_strain}
\end{figure}

\begin{figure}[htbp] \centering
	\subfigure[]{\includegraphics[width = 0.4\textwidth]{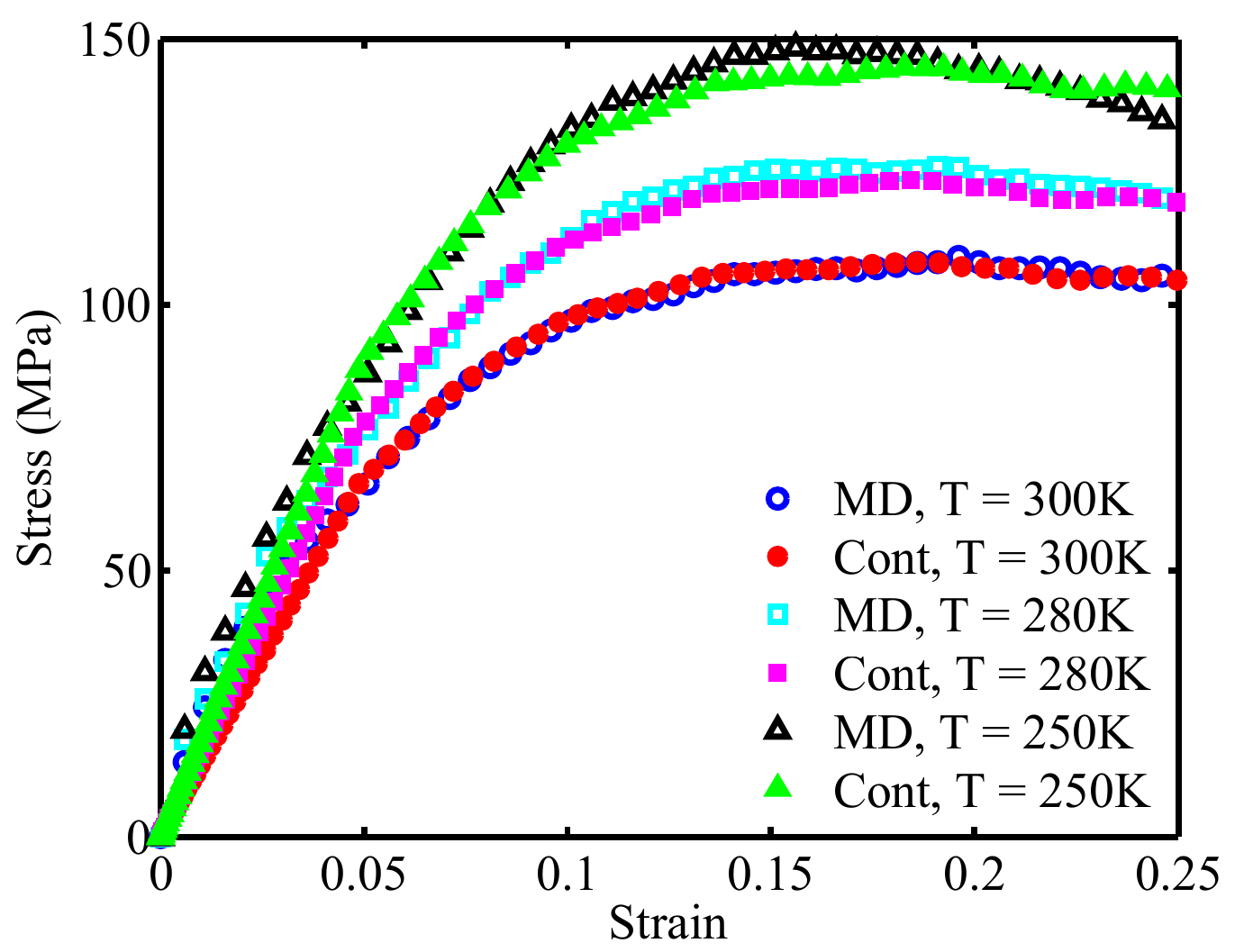}}
	\subfigure[]{\includegraphics[width = 0.4\textwidth]{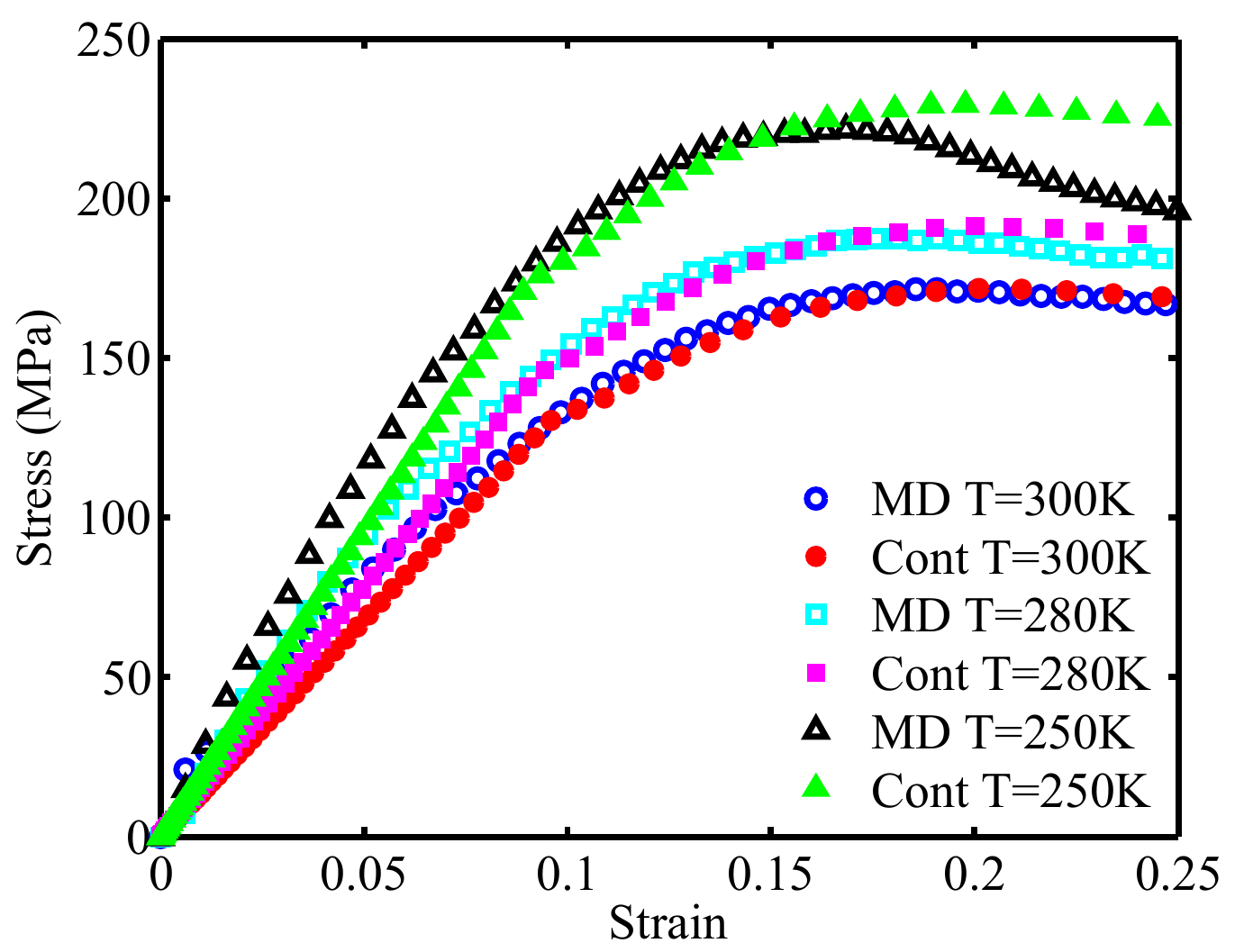}}
	\subfigure[]{\includegraphics[width = 0.4\textwidth]{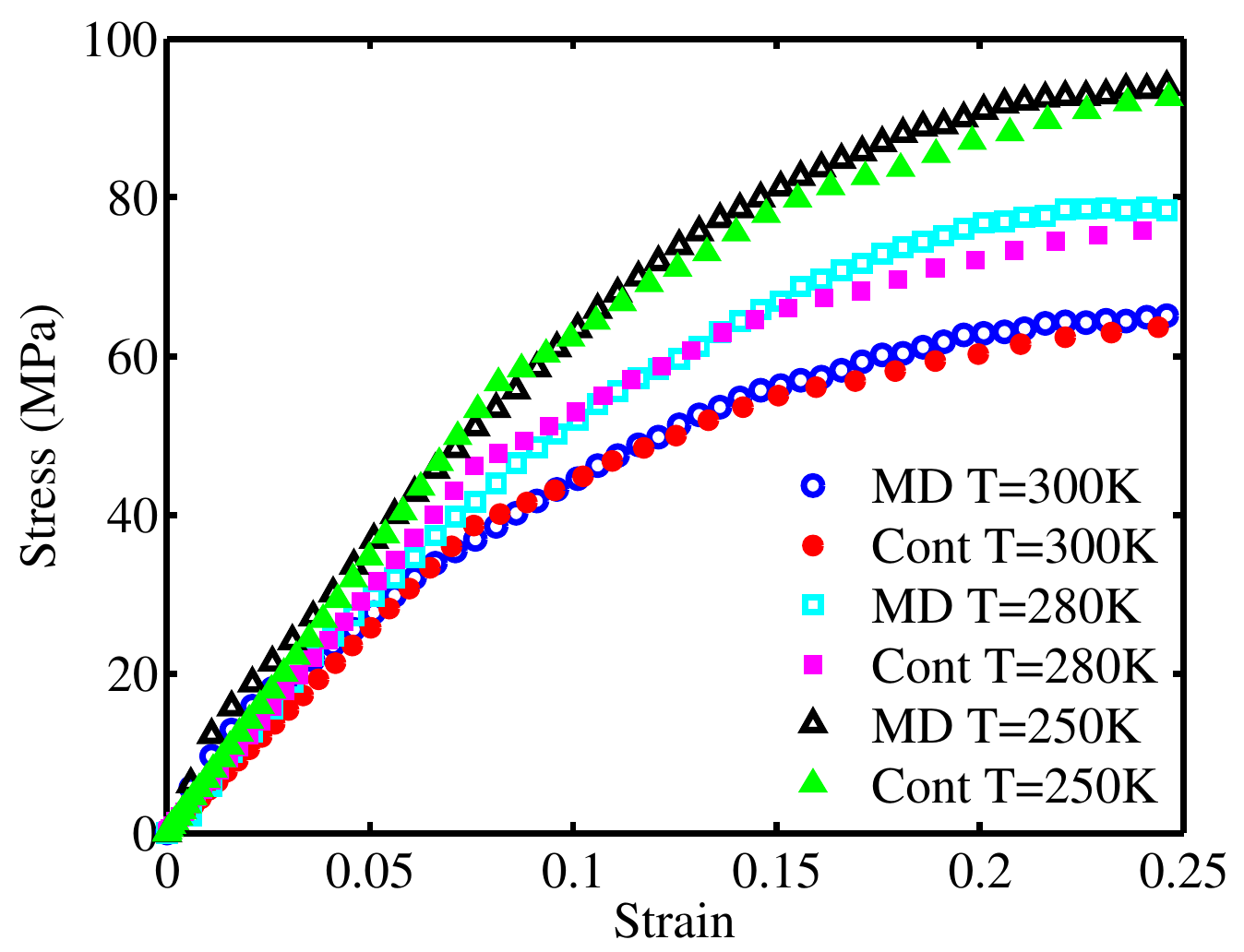}}
	\subfigure[]{\includegraphics[width = 0.4\textwidth]{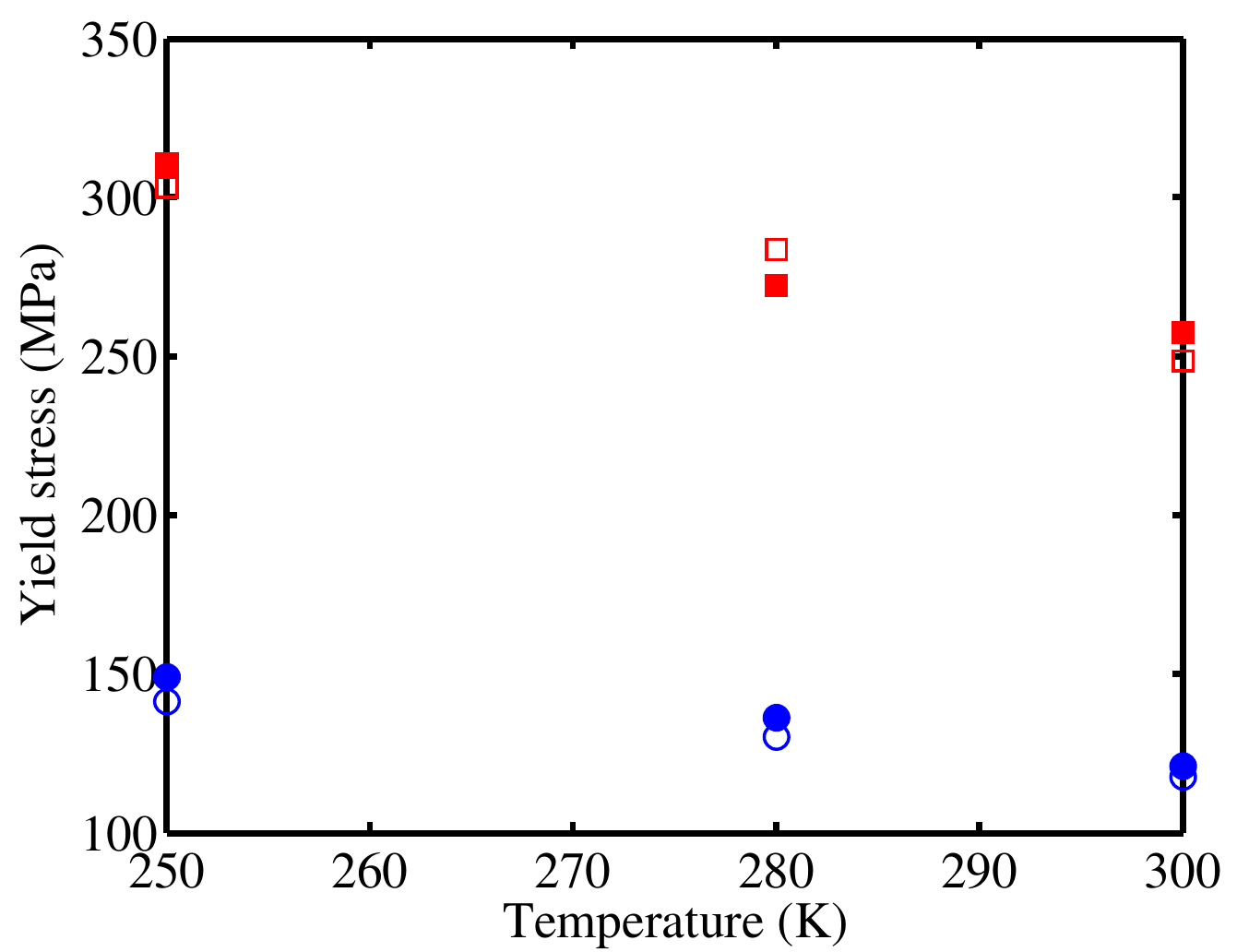}}
	\subfigure[]{\includegraphics[width = 0.4\textwidth]{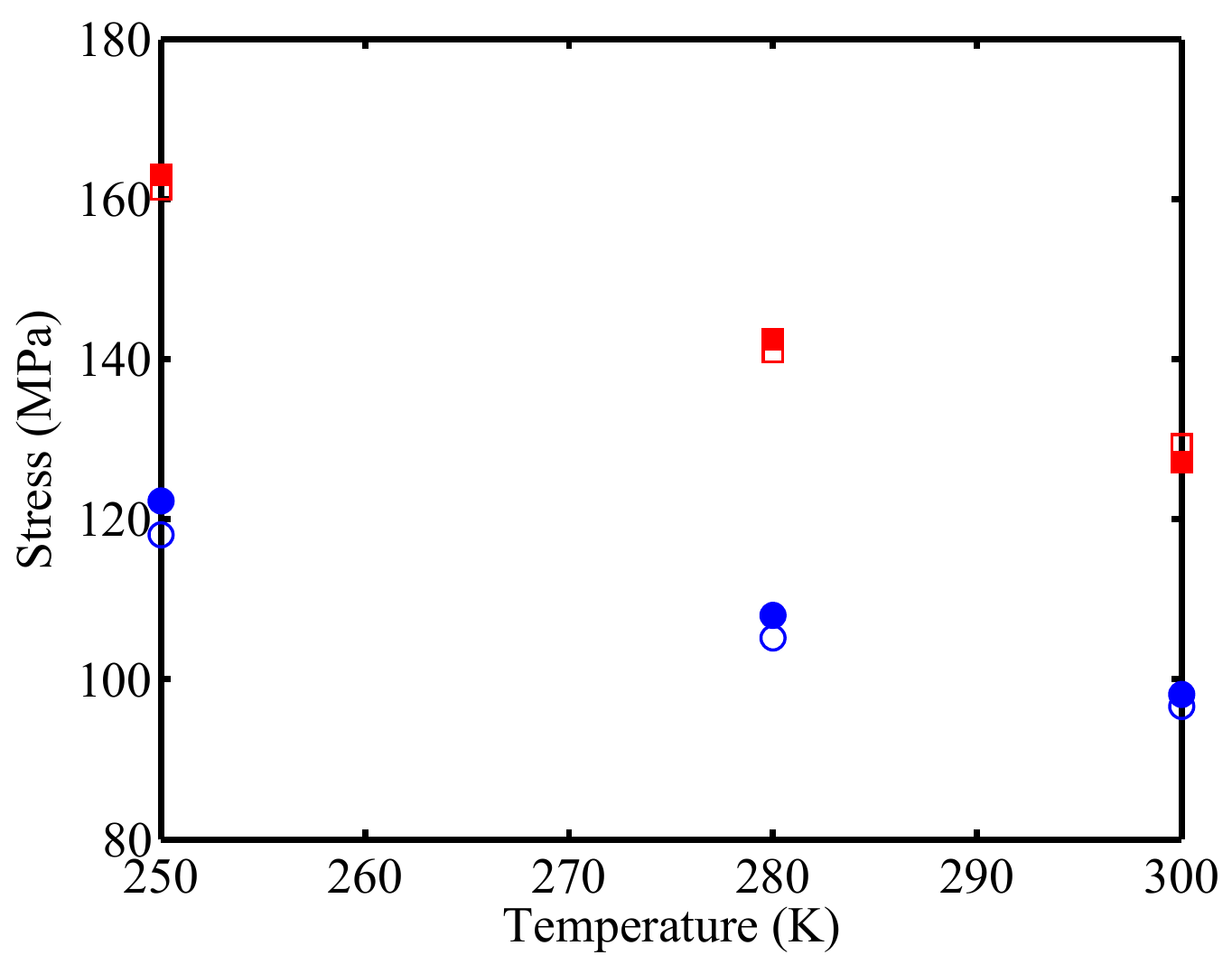}}
	\subfigure[]{\includegraphics[width = 0.4\textwidth]{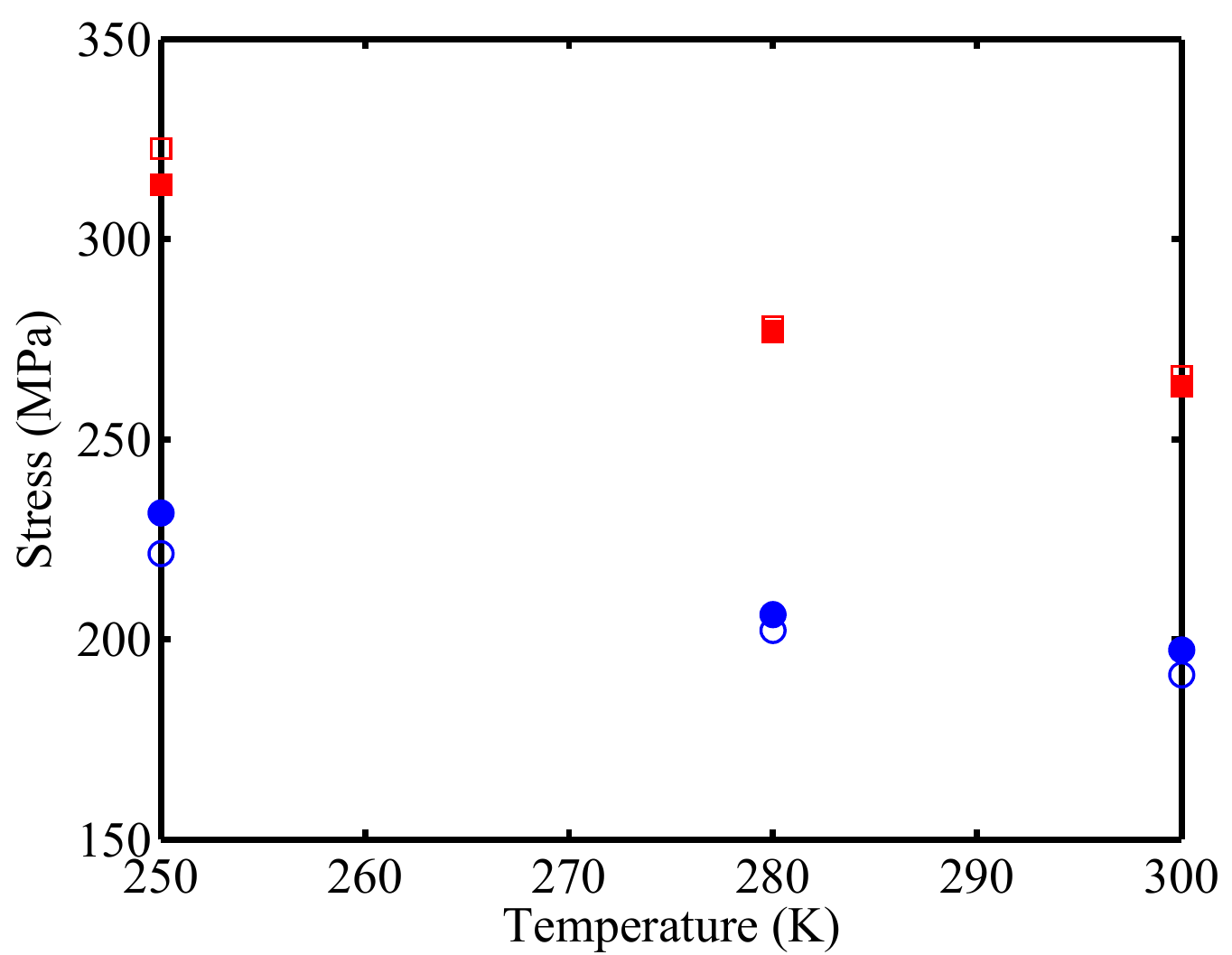}}
	\caption{Comparison of the stress-strain results in (a) tension, (b) compression, (c) shear, (d) equally biaxial tension and compression, the yield stresses (e) unequally biaxial tension ($\dot{\varepsilon}_y = 2 \dot{\varepsilon}_x$) and (f) unequally biaxial compression ($\dot{\varepsilon}_y = 2 \dot{\varepsilon}_x$) predicted by the continuum model and MD simulations at different temperatures. In Figures (d) the tensile and compressive yield stresses obtained from MD simulations are shown by the solid blue circles ({\color{blue}$\bullet$}) and red squares ({\color{red}$\blacksquare$}); the one obtained from the continuum model are shown by hollow blue circles ({\color{blue}$\circ$}) and red squares ({\color{red}$\square$}), In Figures (e) and (f), the solid blue circles and red squares indicate the unequally biaxially tensile and compressive yield stresses obtained from MD simulations and continuum model.}
	\label{fig:verify_tenscomp_stress_strain_difftemps}
\end{figure}

The presented constitutive model in the aforementioned section was implemented as material parameters into ABAQUS to predict the macroscopic stress-strain responses. Comparison between the responses obtained from MD simulations and from the continuum model for different stress states in Figure \ref{fig:verify_stress_strain} shows a good agreement. The temperature dependence of the uniaxial and biaxial tensile, compressive and shear stress-strain responses is also illustrated in Figure \ref{fig:verify_tenscomp_stress_strain_difftemps} and good agreement between the responses obtained from the continuum model and those from MD simulations for different temperatures is observed. Furthermore, the consistency of the stress-strain responses for different stress states (e.g. the unequally biaxial tension and compression with $\dot{\varepsilon}_y = 2 \dot{\varepsilon}_x$ yield stresses) between MD simulations and the continuum model could be expected.

\begin{figure}[htbp] \centering
	\includegraphics[width = 0.5\textwidth]{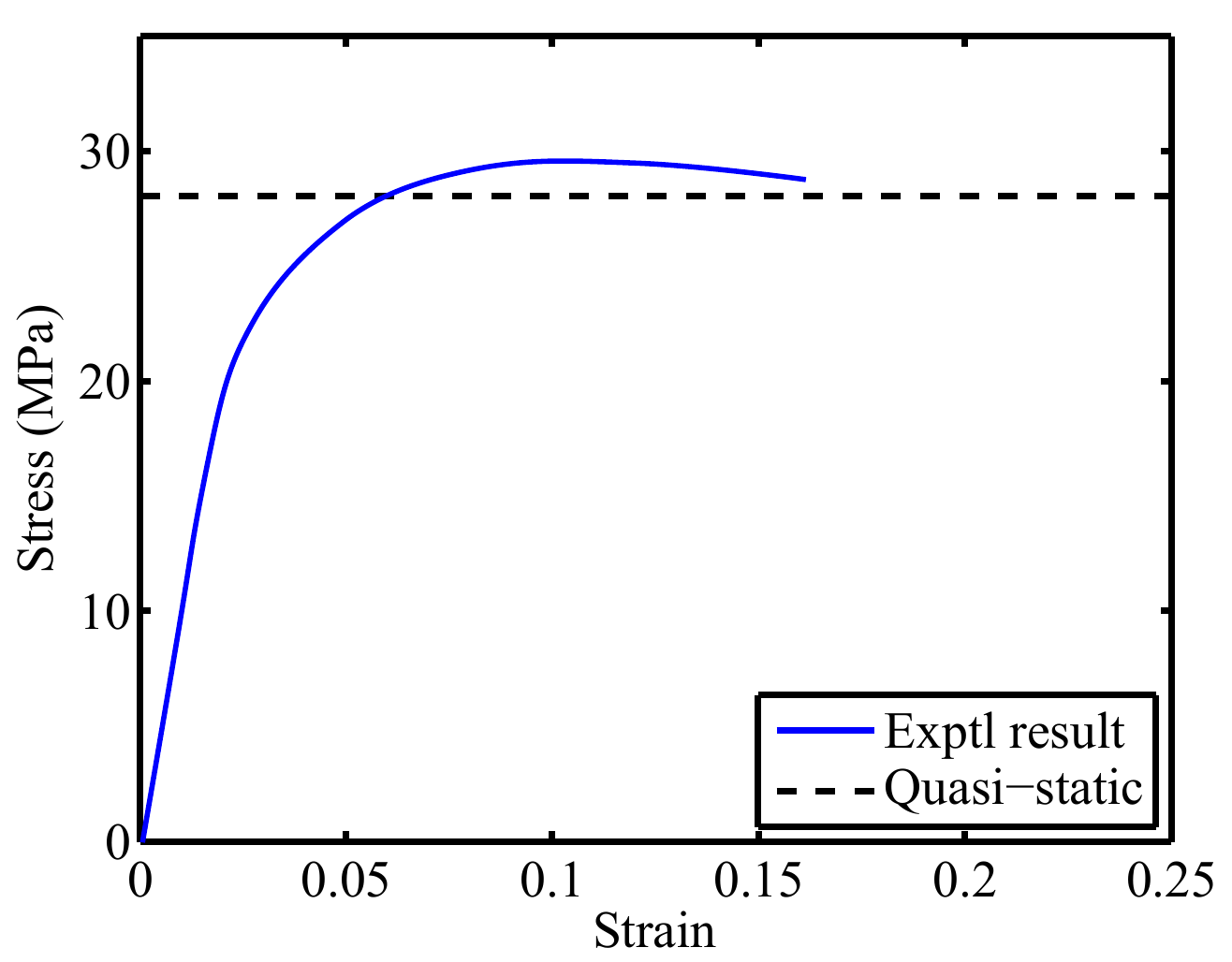}
	\caption{Comparison of the tensile yield stress predicted by the continuum model and experimental result at quasi-static strain rate.}
	\label{fig:compare_stress_strain_with_exptl}
\end{figure}

Figure \ref{fig:compare_stress_strain_with_exptl} compares the predicted yield strength ($\sigma^{quasi}_t = 27.65$ MPa) with the experimental one reported in \cite{Hartmann:1986} ($\sigma^{exptl}_t = 29.3$ MPa) at strain rate of $2 ~\text{min}^{-1}$. Also, the continuum model accurately predicts the compressive yield stress at room temperature, see Table \ref{tab:validate_quasistatic_results}. This proves that the continuum model can be used to accurately predict the yielding occurring at low strain rate.

\section{Conclusions} \label{sec:conclusion}

A hierarchical multiscale model was developed to study the thermo/visco-plastic behavior of the PE. At first, the PLYS and the temperature and strain rate dependent yielding laws were constructed where the constitutive parameters are calibrated from data (yield points for multiaxial stress states) obtained from MD simulations. Then, the scaling law for the entire yield surface was proposed based on the quasi-static tensile simulations at nanoscale.  The yield behavior was upscaled to macroscopic level through an efficient continuum model. The consistency of the results demonstrates that the macroscopic continuum model accurately predicts the behavior achieved from MD simulations.

In addition, validation shows that the tensile and compressive yield stresses are accurately predicted at quasi-static rates by the proposed multiscale multisurface model despite the loss of ad-hoc experimentation. Hence, we believe that this study will open a new door for the design of polymer materials through multiscale simulations, leading to \emph{priori} predictions of yield behavior of polymers.

\section{Acknowledgements}
We gratefully acknowledge the support by ERC COMBAT project (project number 615132).

\end{document}